\documentstyle[aps]{revtex}
\begin{document}
\draft              

\title{Two-hole problem in the t-J model: A canonical transformation 
approach}

\author{V. I. Belinicher, A. L. Chernyshev, and V. A. Shubin}

\address{Institute of Semiconductor Physics, 630090, Novosibirsk, Russia}

\date{\today}
\maketitle

\begin{abstract}
The $t$-$J$ model in the spinless-fermion representation is studied.
An effective Hamiltonian for the quasiparticles is derived using 
canonical transformation approach. It is shown that the rather simple form 
of the transformation generator allows to take into account  
effect of hole interaction with the short-range spin waves and 
to describe the single-hole groundstate. Obtained results are very close to 
ones of the self-consistent Born approximation. 
Further accounting for the long-range spin-wave interaction is possible on 
the perturbative basis. Both spin-wave exchange and an effective interaction 
due to minimization of the number of broken antiferromagnetic bonds are 
included in the effective quasiparticle interaction. Two-hole bound state 
problem is solved using Bethe-Salpeter equation. The only $d$-wave bound 
state is found to exist in the region of $1<(t/J)<5$. Combined effect of 
the pairing interactions of both types is important to its formation. 
Discussion of the possible relation of the obtained results to the problem 
of superconductivity in real systems is presented.
\end{abstract}


\vskip 0.5cm 


\section{Introduction}
\label{sec:level1}

The problem of the hole motion in an antiferromagnetic (AF) background of 
the local spins, originally arisen in connection with the study of the 
localized magnetic insulators, \cite{BR,BNK} has received considerable 
attention since the discovery of the CuO$_2$ based high-temperature 
superconductors.  
It is well established that at zero doping these materials 
are insulators with the long-range AF order, which is well described by the 
two-dimensional Heisenberg model.\cite{Man} 
The instability of long-range AF order under the small finite doping 
of carriers is due to the strong interaction of spins with mobile 
holes. \cite{SS1,SF}
The simplest model, which contains in itself this strong interaction
is the $t$-$J$  model.\cite{And} 
Extensive studies of this 
model validity for the description of the real CuO$_2$ plane result in 
the number of quantitative predictions for the range of parameters and in 
the set of possible $t$-$J$ model 
generalizations.\cite{ZR,Shas,Jeff,Alig,Bel1} 
It is widely believed that the essential low-energy physics of the 
high-T$_c$ systems can be done using the pure $t$-$J$ model 
\begin{eqnarray}
\label{1}
H_{t-J}= t \sum_{\langle ij\rangle,\alpha}
\tilde{c}^{\dag}_{i,\alpha}\tilde{c}_{j,\alpha} + J \sum_{\langle ij\rangle}
\left({\bf S}_i{\bf S}_j - {\textstyle\frac{1}{4}}N_i N_j\right) \ ,  
\end{eqnarray}
in the standard notation of the 
constrained fermion creation (annihilation) operators
$\tilde{c}^{\dag}_{i,\alpha}(\tilde{c}_{i,\alpha})$,
$\langle ij \rangle$ denotes the nearest neighbor sites, ${\bf 
S}_i$ is a local spin operator, $N_i$ is the operator of the number of 
spins. For the consideration of the two- and many-hole problem one must 
include projection operators into $J$-term, which project out the unphysical 
states with both spin and hole at the same site. Such a procedure 
is described in Section V,B.
Physically, $t$-term describes additional hole (singlet) 
hopping over the background of local spins, or, otherwise,
the hopping of hole (vacancy) in the 
spin background. An important feature of this 
term is the absence of the double particle occupancy at site.  
Exclusion of doubly occupied states does not allow to apply
mean field type approximations.

The single-hole problem in the $t$-$J$ model (\ref{1}) has been extensively 
studied by the various analytical 
\cite{KLR,SVR,SS,Trug1,Eder1,Sush1,barab,Becker,horsch,Auer,Li} and 
numerical \cite{Dag1,Elser,Eder2,Good1} techniques, which have provided  the 
deep understanding of the character of the hole motion. For a review see, 
e.g., Refs.  \cite{YuLu,Dag}. Analytical results obtained within the 
self-consistent Born approximation (SCBA) \cite{KLR,SVR,horsch,Man1} 
agree very well with the exact diagonalization studies on clusters, 
\cite{Dag} variational, \cite{Eder1,Sush1,barab} and other approaches. 
\cite{Becker} The main feature of hole motion revealed in these studies 
is the strong renormalization of the naive tight-binding result for the band 
due to hole "dressing" by the cloud of spin excitations, which leads to 
a narrow band ($\sim 2J$ for $t/J>1$ and $\sim t^2/J$ for $t\ll J$) with 
minima at the $\pm (\pi/2, \pm \pi/2)$ points on the boundary of the 
magnetic Brillouin zone (MBZ).  

The two-hole problem has received much attention due to the searching
of possible pairing mechanisms. 
In spite of the large amount of work the full consensus 
on the problem of bound states in the $t$-$J$ model is absent. 
There were a lot of works devoted to the study of the spin-fluctuation 
pairing and corresponding type of superconductivity 
\cite{Schr1,YuLu1,SS2,FKS,Plak}.  There is strong evidence that the 
long-range spin-wave exchange, which is the source of the dipolar 
interaction between holes \cite{SS2,Fren}, can lead to the $d$-wave pairing 
in the $t$-$J$ model.  As it was established in Ref. \cite{SK} the 
corresponding bound states are shallow and have the large size.  Many 
efforts also have been aimed to the study of the $t$-$J$ model bound states 
originated from the fact that the two holes can minimize their energy by 
sharing the common link, that can lead to the picture of superconductivity 
by "preformed" pairs \cite{DagNaz}.  More specifically, numerical works in 
exact diagonalization on small clusters and Monte-Carlo studies, which 
account for the latter interaction, provide negative energy of the bound 
state of the $d_{x^2-y^2}$ symmetry up to the values $t/J\sim 3-5$ 
\cite{Kax,Riera,Poil1,Bonin}, which are relevant to the real compounds.  
Variational \cite{Eder3}, and some kind of quasiparticle calculations 
\cite{CDS} yield the critical value of $t\sim 2J$ for this interaction, 
which is somewhat lower than the real one.  Generally, there is no agreement 
on the energy of the groundstate of two holes and on their spatial 
correlation function \cite{Good2} even between the similar approaches. 

In this paper we propose a canonical transformation approach to the $t$-$J$ 
model problems, that allowed us to turn from the $t$-$J$ model to an 
effective quasiparticle Hamiltonian, describing the "dressed" holes and 
their interaction of the "contact" type and via spin waves, and then to find 
the groundstate of two such quasiparticles. Both types of interactions of 
the most interest are accurately accounted for by our approach.
In some sense, we use the 
ideas of the earlier works by Sushkov {\it et al} 
\cite{Sush1,SF,svert,SK}, where the same scheme was realized using quite 
different approach. 

To begin, let us describe the form of the Hamiltonian (\ref{1}) we start
with. The most popular analytical approach to the $t$-$J$ model is the SCBA 
\cite{KLR,SVR,horsch,Man1}, which is based on the 
spinless-fermion representation for the fermion operators
and Holstein-Primakoff \cite{KLR,horsch} or Dyson-Maleev \cite{Onuf} 
representation for the spin operators for the $t$-$J$ model. 
Namely, this approach is applied to the spin-polaron Hamiltonian, which is 
followed from the $t$-$J$ one (\ref{1}) in the presence of the 
long-range AF order and in the linear spin-wave approximation
\begin{eqnarray}
\label{1a}
H  \simeq 2J\sum_{\bf q}\omega_{\bf q} a_{\bf q}^{\dag}a_{\bf q} 
+ t \sum_{{\bf k},{\bf q}}\left(M_{{\bf 
k},{\bf q}} h_{\bf k-q}^{\dag} h_{\bf k}a_{\bf q}^{\dag} + 
\mbox{H.c.}\right) \ ,
\end{eqnarray}
where $h^{\dag} (h)$, $a^{\dag} (a)$, are the spinless hole and magnon 
operators, respectively,
$2J\omega_{\bf q}=2J(1-\gamma_{\bf q})^{1/2}$ is the spin-wave energy, 
$M_{{\bf k},{\bf q}} = 4 (\gamma_{\bf k-q}u_{\bf q}+\gamma_{\bf k} 
v_{\bf q})$, $u_{\bf q}, v_{\bf q}$ are the Bogolubov canonical  
transformation parameters, $\gamma_{\bf k}=(\cos k_x+\cos k_y)/2$.
The spinless-fermion representation fulfills the above mentioned 
constraint on double occupation exactly \cite{horsch} and, therefore, the 
only approximation made is the spin-wave one. As it was recently shown in 
Ref.  \cite{Man1}, the two-loop corrections due to the higher order terms in 
(\ref{1a}) is analogous to the higher-order nonlinear spin-wave correction 
to the linear spin-wave theory, and have the same order of smallness.

We will study this (\ref{1a}) version of the $t$-$J$ model with 
the additional 
interaction terms from the projection operators in $J$-term. 
In such a formulation (\ref{1a})
the $t$-$J$ problem is explicitly the problem with
the extremely strong interaction. The problem of the interaction of 
fermion excitations with bosonic field and the resulting effective 
"dressing" of fermion by the virtual cloud of bosons is an old and well 
investigated problem, and the powerful approach to it is 
the canonical transformation one \cite{LF}. Therefore, one can hope 
that a canonical transformation will be found very helpful
for the $t$-$J$ model too. Briefly, 
we will show that the rather simple transformation, which takes into 
account the main effect of the strong interaction $\sim t$ and allows to 
consider the rest of the interaction perturbatively, exists.

To complete the consideration of the known facts about the Hamiltonian 
(\ref{1a}) let us note, that in the recent work by Reiter \cite{reiter} an 
exact wave function of the single hole has been obtained within the SCBA 
\begin{eqnarray}
\label{3a}
\tilde{h}_{\bf k}^{\dag}|0\rangle 
=a_{\bf k}&&\biggl[h_{\bf k}^{\dag} + \sum_{\bf q}M_{\bf 
k, q} G_{\bf k-q}(E_{\bf k}-\omega_{\bf q})h_{\bf k-q}^{\dag}
a_{\bf q}^{\dag}+ \dots + \sum_{{\bf q},\dots {\bf q}_{n-1}}
M_{\bf k, q} G_{\bf k-q}(E_{\bf k}-\omega_{\bf q})\times\dots\\
&& \times
M_{{\bf k-q}-\dots-{\bf q}_{n-2},{\bf q}_{n-1}} 
G_{{\bf k-q}-\dots -{\bf q}_{n-1}}(E_{\bf k}-\omega_{\bf q}
-\dots -\omega_{{\bf q}_{n-1}}) h_{{\bf k-q}-\dots -{\bf q}_{n-1}}^{\dag} 
a_{\bf q}^{\dag}\dots a_{{\bf q}_{n-1}}^{\dag} \biggr] 
|0\rangle\ , \nonumber
\end{eqnarray}
where $G_{\bf k}(\omega)$ is an exact hole Green function, $E_{\bf k}$ is 
the hole energy. Since it is an exact wave function of the 
Hamiltonian (\ref{1a}) (within the SCBA) there is no interaction 
of the corresponding quasiparticle with spin-waves, i.e. 
the initially strong interaction is transformed
exactly to the "dressing" of the bare hole, and to an effective 
interaction between such quasiparticles.
Unfortunately, one cannot simply obtain the hole-hole interaction 
extracting $\tilde{h}_{\bf k}^{\dag}$ from the wave function (\ref{3a}) as 
a Fermi operator and averaging $H$ (Eq. (\ref{1a})) over the 
two-hole wave function $\tilde{h}_{\bf k}^{\dag}\tilde{h}_{\bf 
k^{\prime}}^{\dag}|0\rangle$, since $\tilde{h}$, $\tilde{h}^{\dag}$, defined 
by this way, do not obey the usual anticommutation relations.

Briefly, we present an approximate solution of the 
diagonalization problem of the initial Hamiltonian (\ref{1a}). An effective 
Hamiltonian is formulated for the quasiparticles, which have the energy, 
bandwidth, and structure very close to SCBA ones. Further solving 
of the two-hole problem is straightforward. 

Described procedure is valid for the region 
$0<(t/J)<5$. We consider the region $1<(t/J)<5$ as an actual one, since  
considering $t/J$ model as a result of the simple Hubbard or many-band 
Hubbard model mapping, $t/J$ parameter has the lower boundary $t/J\sim 1$ 
below that the mapping procedure is not valid. Moreover, $t/J=5$ corresponds 
to $U/t=20$, which is hardly realized in the real compounds.

The paper is organized as follows. In Section II, we give the comparison of 
the lattice polaron problem with the spin-polaron one and write the 
general form of the transformed $t$-$J$ Hamiltonian. In Sections III, IV, we 
apply the proposed procedure to the Ising case, small $t$ limit, and general 
case of the problem. Section V is devoted to the two-hole problem.

\section{Canonical transformation}

From the formal point of view the spin-polaron Hamiltonian (\ref{1a})
has the form that is very similar to one of the usual lattice polaron 
problem.  We consider here the lattice polaron problem to compare 
these two models in  detail, establish similarities and differences.

The Fr\"{o}hlich Hamiltonian is
\begin{eqnarray}
\label{1C}
H=&&\sum_{\bf k}E_{\bf k}c_{\bf k}^{\dag} c_{\bf k}+ \sum_{\bf q} 
\Omega_{\bf q}b_{\bf q}^{\dag} b_{\bf q}
+ \sum_{{\bf k},{\bf q}} \gamma_{\bf q}\Omega_{\bf q}
c_{\bf k-q}^{\dag} c_{\bf k}(b_{\bf q}^{\dag}+ b_{\bf -q}),
\end{eqnarray}
where $c^{\dag}(c)$ and $b^{\dag}(b)$  are the electron and phonon 
operators, $E_{\bf k}$ and $\Omega_{\bf q}$ are their 
energies, respectively. $\gamma_{\bf q}\Omega_{\bf q}$ is the 
electron-phonon coupling. 
At the limit of the "static" electron ($E_{\bf k}=E_0$) diagonalization 
of the Hamiltonian can be done exactly using the Lang-Firsov 
transformation \cite{LF}:
\begin{eqnarray}
\label{2C}
&&H_{eff}=e^{-S}H e^S =H+[H,S]+\frac{1}{2!}[[H,S]S],
\nonumber \\
&&S= - \sum_{{\bf k},{\bf q}} \gamma_{\bf q}
c_{\bf k-q}^{\dag} c_{\bf k}(b_{\bf q}^{\dag} - b_{\bf -q}) .
\end{eqnarray}
Only the first two commutators are not equal to zero in this limit.
One easily obtains the effective Hamiltonian in the 
terms of the "dressed" electron
\begin{eqnarray}
\label{3C}
H_{eff}=&&\left(E_0-\sum_{\bf q}\Omega_{\bf q}|\gamma_{\bf q}|^2\right)
\sum_{\bf k}c_{\bf k}^{\dag} c_{\bf k}+ 
\sum_{\bf q}\Omega_{\bf q}b_{\bf q}^{\dag} b_{\bf q} 
 - \sum_{{\bf k},{\bf k^{\prime}},{\bf q}} 
\Omega_{\bf q}|\gamma_{\bf q}|^2 c_{\bf k-q}^{\dag} 
c_{\bf k^{\prime}+q}^{\dag} c_{\bf k^{\prime}} c_{\bf k} .  
\end{eqnarray}
Thus the electron-phonon interaction term in Eq. (\ref{1C}) results in the 
lowering of the electron energy (polaronic shift) and the direct $n_i^e 
n_j^e$ interaction.  For the mobile electron an infinite series of terms in 
Eq. (\ref{2C}) arises. It is summed up and yields the effective hopping term 
describing the collective hopping process of the bare electron with the 
cloud of phonons.  It was shown that this "dressing" leads to the 
exponentially narrow effective band \cite{LF}.  The emitting of the phonons 
(multiple phonon processes) can be considered as the perturbation.  In the 
basis of this approach there is a clear physical idea that the presence of 
the electron at the lattice site leads to the change of the
equilibrium positions of the surrounding ions, and the new 
eigenfunction of phonons is a coherent state. 

There are two main differences between the phonon and magnetic 
polaron problems.  The first one is the principal absence of the 
"bare" dispersion in the Hamiltonian (\ref{1a}), i.e. its 
hopping term is more likely the vertex than the usual 
tight-binding hopping integral \cite{YuLu}. The second one is 
the nonlocal character of the hole-spin interaction, i.e. 
emission (absorption) of magnon can be done only by hopping. 
Because of that there is no "static" limit of the problem even 
if $t\ll J$, and the evident {\it a priori} ideas about the 
structure of spin cloud around the hole are absent.

Nevertheless, the existing knowledge of the character of hole motion in 
AF background can help one to succeed in turning to an effective 
model, which is much more appropriate to study than the initial
one. Firstly, at the Ising background the groundstate of the hole is the 
localized magnetic polaron, which is formed  by self-retraceable motion 
of the hole. At the N\'{e}el background  there is the  
similar situation, i.e. spin waves in the virtual spin cloud around 
the hole are absorbed exactly in the reversed order than they were emitted. 
The contribution of the processes beyond these retraceable paths (or SCBA) 
 approximation is found to be extraordinary small. 
Secondly, it was found in the number of works that taking into account 
the hole "dressing" even by the single spin wave already provides results, 
which are close to the exact ones \cite{Sush1}, \cite{barab}. 
Namely, the bottom of the band, hole minima locations, and width of the 
 band are determined with a sufficient accuracy \cite{Sush1}. Therefore, 
it shows that the main contribution to the polaron well 
formation for the actual range of $(t/J)<5$ is made by the 
"one-string" component of the hole wave-function (\ref{3a}).  
 Authors of some SCBA works also successfully used this approximation 
for some other $t$-$J$ model studies \cite{reiter}, \cite{rhorsch}.  These 
are the reasons to hope that relatively simple transformation in the 
spirit of Lang-Firsov one can be used to get the effective model, which 
accounts for the main polaron effect (of the order of $t$) in the hole 
 energy and hole-hole interaction, whereas the other included terms allow 
to apply the perturbation theory.

We propose the general form of the generator of such a transformation:
\begin{eqnarray}
\label{4C}
 S= \sum_{{\bf k},{\bf q}} \mu_{{\bf k},{\bf 
q}}\left(h_{\bf 
k-q}^{\dag} h_{\bf k} a_{\bf q}^{\dag} - \mbox{H.c.}\right),
\end{eqnarray}
where $\mu_{{\bf k},{\bf q}}$  is the parameter of the transformation.
It is natural to require for $\mu_{{\bf k},{\bf q}}$ to obey 
the same symmetry properties as the kinematic factor $M_{{\bf k},{\bf q}}$
of the $t$-term in the Hamiltonian (\ref{1a}). Note that $M_{{\bf k},{\bf 
q}}$ is odd with respect to the transformations  $M_{{\bf k},{\bf 
q}}=-M_{{\bf k}+{\bf Q},{\bf q}}=-M_{{\bf k},{\bf q} +{\bf Q}}$, here ${\bf 
Q}=(\pi,\pi)$. So, without loss of generality one can rewrite $\mu_{{\bf 
k},{\bf q}}= f_{\bf k,q} M_{{\bf k},{\bf q}}$, where $f_{{\bf k},{\bf q}}$ 
is even under mentioned symmetry transformations.

Main terms of the spin-hole Hamiltonian (\ref{1a}), which 
 we wish to decouple are:
\begin{eqnarray}
\label{5C}
H=t \sum_{{\bf k},{\bf q}} M_{{\bf k},{\bf 
q}}\left(h_{\bf 
k-q}^{\dag} h_{\bf k} a_{\bf q}^{\dag} + \mbox{H.c.}\right) + 
2J\sum_{\bf q}\omega_{\bf q} a_{\bf q}^{\dag} a_{\bf q},
\end{eqnarray}
after the applying of the transformation (\ref{4C}) they provide transformed
Hamiltonian
\begin{eqnarray}
\label{6C}
H_{tr}=&&\sum_{\bf k}E_{\bf k}
h_{\bf k}^{\dag} h_{\bf k}+ 2J\sum_{\bf q} \omega_{\bf q}
a_{\bf q}^{\dag} a_{\bf q} 
+ t
 \sum_{{\bf k},{\bf k^{\prime}},{\bf q}} V^{hh}_{{\bf k},{\bf k^{\prime}},{\bf q}} 
h_{\bf k-q}^{\dag} h_{\bf k^{\prime}+q}^{\dag} h_{\bf k^{\prime}} h_{\bf k} 
+ t \sum_{{\bf k},{\bf q}}F_{\bf {\bf k},{\bf q}} 
M_{{\bf k},{\bf q}}\left(h_{\bf k-q}^{\dag} h_{\bf k} 
a_{\bf q}^{\dag} + \mbox{H.c.}\right) 
\nonumber\\
&&
+ t \sum_{{\bf k},{\bf q},{\bf q^{\prime}}} V^{haa}_1
({\bf k},{\bf q},{\bf q^{\prime}})
\left(h_{\bf k-q+q^{\prime}}^{\dag}
 h_{\bf k} a_{\bf q}^{\dag} a_{\bf q^{\prime}}^{\dag}
+ \mbox{H.c.}\right)  
+ t \sum_{{\bf k},{\bf q},{\bf q^{\prime}}} V^{haa}_2
({\bf k},{\bf q},{\bf q^{\prime}})
h_{\bf k-q+q^{\prime}}^{\dag} h_{\bf k} 
a_{\bf q}^{\dag} a_{\bf q^{\prime}} + \dots \ ,
\end{eqnarray}
where we omit terms with six and more fermion operators.
General expressions for the hole energy $E_{\bf k}$, hole-magnon formfactor
$F_{\bf k,q}$ (up to the sixth order of the transformation), hole-hole vertex
$V^{hh}_{{\bf k},{\bf k^{\prime}},{\bf q}}$ (up 
to the fourth order) and other vertices are given in Appendix A.
There is the freedom in the choosing of the parameter of the transformation 
$f_{{\bf k},{\bf q}}$.
By analogy with an exact single-hole wave function (\ref{3a}) one can 
propose the following form of the parameter
$f_{{\bf k},{\bf q}}=f_{\bf k}G({\bf k-q},E_{\bf k}-2J\omega_{\bf q})$, 
where $G({\bf k},\omega)$ is the Green function of the dressed hole.
As a result one obtains the very complex 
self-consistent integral equation on $E_{\bf k}$ and self-energy 
$\Sigma_{\bf k}(\omega)$, which is hardly soluble.
There are some other ways of choosing the transformation parameter (TP),
which allow to avoid self-consistency in equations and technically are 
more advantageous. The simplest one is to neglect ${\bf q}$-dependence in 
$f_{\bf k,q}\Rightarrow f_{\bf k}$, and
then to determine it as a solution of an equation obtained
from the requirement of the minimum of the hole energy, or, for example, 
equality to zero of the hole-magnon formfactor.
We will discuss this approach in the next two 
Sections. Here we claim that for the rather general form of TP
one can restrict oneself by the first four terms in the 
transformed Hamiltonian (\ref{6C}), namely 
\begin{eqnarray}
\label{8C}
H_{eff}=\sum_{\bf k}E_{\bf k}
h_{\bf k}^{\dag} h_{\bf k}+ 2J\sum_{\bf q} \omega_{\bf q}
a_{\bf q}^{\dag} a_{\bf q} 
+ t \sum_{{\bf k},{\bf k^{\prime}},{\bf q}} 
V^{hh}_{{\bf k},{\bf k^{\prime}},{\bf q}} 
h_{\bf k-q}^{\dag} h_{\bf k^{\prime}+q}^{\dag} h_{\bf k^{\prime}} h_{\bf k} 
+ t \sum_{{\bf k},{\bf q}}F_{\bf {\bf k},{\bf q}} 
M_{{\bf k},{\bf q}}\left(h_{\bf k-q}^{\dag} h_{\bf k} 
a_{\bf q}^{\dag} + \mbox{H.c.}\right) 
\end{eqnarray}
keeping in mind that the  transformed "additional" interactions ($J$-term)  
is included in $V^{hh}_{{\bf k},{\bf k^{\prime}},{\bf q}}$.  Moreover, 
resulting effective hole-magnon vertex is
perturbative, i.e. its second-order addition to the energy 
$\delta E_{\bf k}$ is small.  Importance of this vertex for the two-hole 
problem we will discuss in the Section V.

\section{Ising limit}

Let us start the general consideration of our approach from the Ising 
case.  As it was noted in Ref. \cite{star} treating the $t$-$J$ model
 in the Ising limit within 
the linear spin-wave approximation remains the physics 
of the problem essentially unchanged. Moreover, it was shown \cite{star} 
that the spin-wave formalism provides exactly the same result as one of 
the SCBA.

 One can get the Ising limit of the $t$ and $J$ terms 
from the general spin-hole Hamiltonian Eq. (\ref{1a}) 
using the momentum-independence of $\omega_{\bf 
q}$ at the Ising background and the absence of the spin fluctuations 
($u_{\bf q}=1$ and $v_{\bf q}=0$):
\begin{eqnarray}
\label{1I}
H= &&t
\sum_{{\bf k},{\bf q}} M_{{\bf k},{\bf q}}^I\left(h_{\bf 
k-q}^{\dag} h_{\bf k} a_{\bf q}^{\dag} + \mbox{H.c.}\right) + 
2J\sum_{\bf q} a_{\bf q}^{\dag} a_{\bf q},\\&& \mbox{with} \ \ \ \ M_{{\bf 
k},{\bf q}}^I = 4 \gamma_{\bf k-q} ,\nonumber 
\end{eqnarray}
 the additional terms
 of the interaction Hamiltonian (from $J$-term) can 
be considered independently and we will include them later.

Following the analogy with the Lang-Firsov transformation we turn to the 
effective Hamiltonian with the help of the transformation:
\begin{eqnarray}
\label{2I}
&&H_{eff}=e^{-S}H e^S =H+[H,S]+\frac{1}{2!}[[H,S]S]+\dots,
\nonumber \\
&&S= f 
\sum_{{\bf k},{\bf q}} M_{{\bf k},{\bf q}}^I\left(h_{\bf 
k-q}^{\dag} h_{\bf k} a_{\bf q}^{\dag} - \mbox{H.c.}\right),
\end{eqnarray}
where the generator of the transformation simply reproduces the kinematic 
structure 
of the hopping Hamiltonian and involves the single free parameter $f$.
It is natural  for parameter of the transformation 
$f$ to be {\bf k}-independent, since the energy of the hole  for the Ising 
problem does not depend on ${\bf k}$. 
Using 
the evident relation $[H_J,S]=f(2J/t)H_t$ 
one can make the effective Hamiltonian calculation and get:
\begin{eqnarray}
\label{3I}
H_{eff}=&&E_h\sum_{\bf k}h_{\bf k}^{\dag} h_{\bf k}+ 2J\sum_{\bf q} 
a_{\bf q}^{\dag} a_{\bf q} 
+ t
 \sum_{{\bf k},{\bf k^{\prime}},{\bf q}} 
V^{hh}_{{\bf k},{\bf k^{\prime}},{\bf q}} 
h_{\bf k-q}^{\dag} h_{\bf k^{\prime}+q}^{\dag} 
h_{\bf k^{\prime}} h_{\bf k} \nonumber\\
&&+ 
t F
\sum_{{\bf k},{\bf q}} M_{{\bf k},{\bf q}}^I\left(h_{\bf 
k-q}^{\dag} h_{\bf k} a_{\bf q}^{\dag} + \mbox{H.c.}\right) 
+ t
 \sum_{{\bf k},{\bf q},{\bf q^{\prime}}} 
V^{haa}_{{\bf k},{\bf q},{\bf q^{\prime}}} 
h_{\bf k-q+q^{\prime}}^{\dag} h_{\bf k} 
a_{\bf q}^{\dag} a_{\bf q^{\prime}} + \dots 
\end{eqnarray}
with the one-hole energy, hole-magnon formfactor, hole-hole vertex, and 
hole-two magnon vertex:
\begin{eqnarray}
\label{4I}
E_h\ \ \ \simeq&&8t\left(f-{\textstyle\frac{4}{3}}f^3
+\frac{2J}{t}\left({\textstyle\frac{1}{2}}f^2 
-{\textstyle\frac{1}{3}}f^4
\right)\right)\nonumber\\
F\ \ \ \ \simeq&&1-4f^2+
\frac{2J}{t}\left(f-{\textstyle\frac{4}{3}}f^3
\right)
\\
V^{hh}_{{\bf k},{\bf k^{\prime}},{\bf q}}=&&\left(M_{{\bf k},{\bf q}}^I
M_{{\bf k^{\prime}+q},{\bf q}}^I+M_{{\bf k-q},{\bf -q}}^I
M_{{\bf k^{\prime}},{\bf -q}}^I\right)
\cdot 
f \left[1+\frac{J}{t}f-{\textstyle\frac{4}{3}}f^2\left(1+\frac{J}{2t}f 
\right)\left(4+\gamma_{\bf k+k^{\prime}} \right)\right]- \nonumber\\ 
&&-
\left(M_{{\bf k},{\bf q}}^I M_{{\bf k-q},{\bf -q}}^I+
M_{{\bf k^{\prime}+q},{\bf 
q}}^I M_{{\bf k^{\prime}},{\bf -q}}^I\right)
\cdot{\textstyle\frac{8}{3}}f^3\left(1+\frac{J}{2t}f \right)
\nonumber\\
V^{haa}_{{\bf k},{\bf q},{\bf q^{\prime}}}=&&
\left(M_{{\bf k},{\bf q}}^I M_{{\bf 
k-q+q^{\prime}},{\bf q^{\prime}}}^I-
M_{{\bf k-q^{\prime}},{\bf -q^{\prime}}}^I M_{{\bf k-q^{\prime}},{\bf 
q}}^I\right) \cdot 2f^2  \left(1+\frac{J}{t}f\right) \nonumber 
\end{eqnarray}
up to the fourth order of transformation for the hole 
energy and formfactor of the spin-hole vertex and 
for the hole-hole interaction and higher vertices.
The first peculiar feature of the Ising case appeares here. 
Minimization of the energy provides an equation on $f$: 
\begin{eqnarray}
\label{4Ia}
&&\frac{\delta E_h}{\delta f} \sim 1-4f^2+
(2J/t)\left(f-{\textstyle\frac{4}{3}}f^3
\right)=0 
\end{eqnarray}
that coincides with equality to zero of the hole-magnon formfactor $F$. 
This is closely connected to the fact that the each act of 
emission or absorption of magnon
is due to the hole hopping, and the polaron is created by moving
on the self-retraceable paths. The role of the so called Trugman processes 
\cite{Trug1} among the other fifth and sixth order contributions 
was found small ($\sim (1/8)$ of relative magnitude) and they 
remain all results essentially
unchanged. The next simplifying
fact is the absence of the two-magnon vertices with the 
$h^{\dag}h a^{\dag}a^{\dag}$ ($aa$) terms. It means that at zero 
temperature there are no contributions of the hole-two-magnon part 
into the self-energy and to the hole-hole vertex, and hence, the 
$h^{\dag}h a^{\dag}a$
 can be omitted in the effective 
model. Thus, after energy minimization the effective Hamiltonian has 
the form, which is very similar to the lattice polaron one:
\begin{eqnarray}
\label{5I}
H_{eff}=&&E_h\sum_{\bf k}h_{\bf k}^{\dag} h_{\bf k}+ 2J\sum_{\bf q} 
a_{\bf q}^{\dag} a_{\bf q}
+ t
\sum_{{\bf k},{\bf k^{\prime}},{\bf q}} 
V^{hh}_{{\bf k},{\bf k^{\prime}},{\bf q}} 
h_{\bf k-q}^{\dag} h_{\bf k^{\prime}+q}^{\dag} 
h_{\bf k^{\prime}} h_{\bf k} \ ,
\end{eqnarray}
here the energy and hole-hole vertex are given by Eq. (\ref{4I}) with $f$ 
obtained from Eq. (\ref{4Ia}).

Eqs. (\ref{4I}), (\ref{4Ia}) for the hole energy show that:
\begin{eqnarray}
\label{8I}
&& f=-\frac{t}{2J}, \ \ \hskip 1.cm \ t/J \rightarrow 0
\nonumber \\
&& f^2\simeq\frac{1}{z}\left(1-\frac{2J}{\sqrt{z}t}\right)
, \ \ \ t/J \gg 1 
\end{eqnarray}
demonstrating the perturbative nature of our approach at small $t/J$ and 
some kind of $1/z$ expansion at large $t/J$.

An exact result for the single-hole energy on the Ising background 
was obtained in Ref. \cite{star} in the form of the difference 
equation. Also, there is an analytical solution of this equation in the 
$t/J\gg 1$ limit first proposed by Bulaevskii, Nagaev, and Khomskii 
\cite{BNK}: 
\begin{eqnarray}
\label{6I}
E = -2\sqrt{z}t - 2J + 2.34 (2J)^{2/3}(\sqrt{z}t)^{1/3}.
\end{eqnarray}
Figure \ref{fig1} presents numerical solution of the exact equation 
\cite{star} (bold solid curve) and approximate solution (\ref{6I}) (dashed 
curve) together with our results Eq. (\ref{4I}). Upper and lower curves 
correspond to the hole energy calculated up to the fourth
and sixth order of the transformation, respectively.
Figure \ref{fig2} shows the components of magnetic polaron wave function
for the exact solution (bold curves)
and ones of our canonically transformed polaron:
\begin{eqnarray}
\label{7I}
\tilde{h^{\dag}}|0\rangle =e^{-S} h^{\dag}|0\rangle
\end{eqnarray}
These figures demonstrate that our single-hole energy is very close to the 
exact one, the higher orders for the actual $t/J$ region play the role of 
corrections, and the components of exact and our polaron wave 
functions are close to each other.  Therefore, one can hope that 
the consideration of the interaction of our quasiparticles will reveal the 
properties of exact eigenstates of the $t$-$J$ Hamiltonian. 

\section{N\'{e}el case.}
\subsection{ Small $t$ limit.}

Let us consider first the small $t/J$ limit of the model. Perturbation 
theory over this parameter works very well and the most results can be 
obtained analytically. Our transformation procedure in this limit also has a 
perturbative sense and it is possible to compare our results with ones of 
the usual perturbation theory. It is also useful to consider this limit for 
the demonstration of some details of our approach.

As it was mentioned in Section II, there are some evident forms of the 
TP $f_{\bf k, q}$. The first one we wish to 
study is $f_{\bf 
k, q}\Rightarrow f_{\bf k}$.  The second one is $f_{\bf k, q}\Rightarrow 
f_{\bf k}/\omega_{\bf q}$, which can be considered as the simplification of 
one proposed in Section II $f_{\bf k, q}\Rightarrow f_{\bf k} G({\bf 
k}-{\bf q}, E_{\bf k}-2J\omega_{\bf q})$ in the small $t$ limit when 
$E_{\bf k}\ll 2J\omega_{\bf q}$. Technical advantage of both cases for any
$t$  is that ${\bf k}$ and ${\bf q}$ dependent parts are 
separable and all arisen integrals can be  reduced to the several 
functions. 

It is interesting to clarify the physical meaning of both transformations. 
Let us show the polaron wave functions:
\begin{eqnarray}
\label{N1}
\mbox{1st case}\hskip 1.cm &&\tilde{h_{\bf k}^{\dag}}|0\rangle 
=\left[\left(1-{\textstyle\frac{1}{2}}f_{\bf k}^2\sum_{\bf q}
M_{\bf k, q}\right)
 h_{\bf k}^{\dag} + f_{\bf k}\sum_{\bf q}M_{\bf 
k, q} h_{\bf k-q}^{\dag}a_{\bf q}^{\dag}+ 
O((t/J)^3)\right]|0\rangle \nonumber\\
\mbox{2nd case}\hskip 1.cm &&\tilde{h_{\bf k}^{\dag}}|0\rangle 
=\left[\left(1-{\textstyle\frac{1}{2}}f_{\bf k}^2\sum_{\bf q}
\frac{M_{\bf k, q}^2}{\omega_{\bf q}}\right)
 h_{\bf k}^{\dag} + f_{\bf k}\sum_{\bf q}\frac{M_{\bf 
k, q}}{\omega_{\bf q}} h_{\bf k-q}^{\dag}a_{\bf q}^{\dag}+ 
O((t/J)^3)\right]|0\rangle \ .
\end{eqnarray}
Since at ${\bf q}\rightarrow 0$ $M_{\bf k, q}\rightarrow 0$ the admixture 
of the long-range magnons in the wave function of the "first case" polaron
is small, i.e. this transformation corresponds to taking into account 
short-range spin-wave "dressing" of the hole. Comparing the Reiter's wave 
function (\ref{3a}) with the "second-case" one suggests their identity for 
this limit. 

Transformed Hamiltonian (\ref{6C}) for the small $t$ limit takes the form:
\begin{eqnarray}
\label{N2}
H_{eff}=&&\sum_{\bf k} \left[\sum_{\bf q}f_{\bf k,q} M_{\bf k,q}^2
\left(1+\frac{J}{t} f_{\bf k,q}\omega_{\bf q}\right)\right]
h_{\bf k}^{\dag} h_{\bf k}+ 2J\sum_{\bf q} \omega_{\bf q}
a_{\bf q}^{\dag} a_{\bf q}\nonumber \\
&&+ t \sum_{{\bf k},{\bf q}} 
M_{{\bf k},{\bf q}}\left(1+\frac{2J}{t} f_{\bf k,q}\omega_{\bf q}\right)
\left(h_{\bf k-q}^{\dag} h_{\bf k} 
a_{\bf q}^{\dag} + \mbox{H.c.}\right) \\
&&+ t \sum_{{\bf k},{\bf k^{\prime}},{\bf q}} 
M_{\bf k, q} M_{\bf k^{\prime}+q, q} f_{\bf 
k^{\prime}+q, q} \left(1+\frac{J}{t} f_{\bf k,q}\omega_{\bf q}\right)
\left(h_{\bf k-q}^{\dag} h_{\bf k^{\prime}+q}^{\dag} 
h_{\bf k^{\prime}} h_{\bf k} + 
\mbox{H.c.}\right) + O(\frac{t^3}{J^3})\ ,
\nonumber
\end{eqnarray}
here we omit hole-two-magnon vertices ($O(t^2/J^2)$) since they 
contribute in the energy and scattering amplitude only in the second 
order ($O(t^4/J^4)$). The hole energy for both types of 
the transformation:  
\begin{eqnarray} 
\label{N3} 
E_{\bf k}= \left\{ 
\begin{array}{ll}
{\displaystyle I_{\bf k} f_{\bf k}+\frac{J}{t}\widetilde{I}_{\bf k} f_{\bf 
k}^2}
& {\displaystyle\mbox{1st case}}\\
{\displaystyle
I_{\bf k}^{\omega} f_{\bf k}\left(1+\frac{J}{t}f_{\bf k}\right)}
&{\displaystyle\mbox{2nd case}}           \ ,
\end{array}
\right.
\end{eqnarray}
with useful notations $I_{\bf k}=\sum_{\bf q}M_{\bf k,q}^2$, 
$\widetilde{I}_{\bf k}=\sum_{\bf q}M_{\bf k,q}^2\omega_{\bf q}$, 
$I_{\bf k}^{\omega}=\sum_{\bf q}M_{\bf k,q}^2/\omega_{\bf q}$.
Minimization of the energy averaging over the band
 provides the following parameters of the 
transformations and formfactors:
\begin{eqnarray}
\label{N4}
\frac{\delta }{\delta f_{\bf k}}\left(
 \sum_{\bf k^{\prime}} E_{\bf k^{\prime}}\right)=0 \Rightarrow 
\left\{
\begin{array}{l}
{\displaystyle f_{\bf k}=-\frac{t}{2J}\frac{I_{\bf k}}
{\widetilde{I}_{\bf k}}}               \\
{\displaystyle f_{\bf k}=-\frac{t}{2J}}
\end{array}
\right. \hskip 1.cm
F_{\bf k,q}=
\left\{
\begin{array}[c]{ll}
{\displaystyle 1-\omega_{\bf q}\frac{I_{\bf k}}{\widetilde{I}_{\bf k}}}
&{\displaystyle\mbox{1st case}}  \\
{\displaystyle  0}
&{\displaystyle\mbox{2nd case}}
\end{array}
\right.                                      \ ,
\end{eqnarray}
and the effective Hamiltonians
\begin{eqnarray}
\label{N5}
H_{eff}^{1}=&&-\frac{t^2}{2J}\sum_{\bf k}\frac{I_{\bf k}^2}
{\widetilde{I}_{\bf k}}h_{\bf k}^{\dag} h_{\bf k}+ 2J\sum_{\bf q} 
\omega_{\bf q} a_{\bf q}^{\dag} a_{\bf q}
+t\sum_{\bf k, q} 
M_{\bf k, q} \left(1-\omega_{\bf q}\frac{I_{\bf k}}
{\widetilde{I}_{\bf k}}\right) 
\left(h_{\bf k-q}^{\dag} h_{\bf k} a_{\bf q}^{\dag} + 
\mbox{H.c.}\right) 
\nonumber\\
&&-\frac{t^2}{2J}\sum_{\bf k, k^{\prime}, q} 
M_{\bf k, q} M_{\bf k^{\prime}+q, q}\frac{I_{\bf k^{\prime}+q}}
{\widetilde{I}_{\bf k^{\prime}+q}}\left(1-\omega_{\bf q}\frac{I_{\bf k}}
{2\widetilde{I}_{\bf k}}\right) 
\left(h_{\bf 
k-q}^{\dag} h_{\bf k^{\prime}+q}^{\dag} 
h_{\bf k^{\prime}} h_{\bf k} + \mbox{H.c.}\right) 
\\
H_{eff}^{2}=&&-\frac{t^2}{2J}\sum_{\bf k}I_{\bf k}^{\omega}
h_{\bf k}^{\dag} h_{\bf k}+ 2J\sum_{\bf q} 
\omega_{\bf q} a_{\bf q}^{\dag} a_{\bf q}
-\frac{t^2}{2J}\sum_{\bf k, k^{\prime}, q} 
\frac{M_{\bf k, q} M_{\bf k^{\prime}+q, q}}{\omega_{\bf q}}
\left(h_{\bf 
k-q}^{\dag} h_{\bf k^{\prime}+q}^{\dag} h_{\bf k^{\prime}} 
h_{\bf k} + \mbox{H.c.}\right)                  \ .
\nonumber
\end{eqnarray}
The hole energies have the well known shape with the 
minima at $\pm(\pi/2,\pm\pi/2)$ points and large effective mass along 
MBZ boundary.
Note, that the "second-case" transformation, which is the Lang-Firsov one
by the construction, diagonalizes starting Hamiltonian exactly. 
Evidently, the "second-case" results for the hole energy and hole-hole 
vertex coincide with these values obtained by the usual perturbation theory.  
The first case correspond to the separation of the scales and
implies the perturbative account for the interaction with the long-range 
spin waves.  Clearly, that accounting for the second-order contribution 
of the remaindering part of the hole-magnon vertex ($H_{eff}^{1}$ 
(\ref{N5})) into the hole energy and hole-hole vertex reproduces results of 
the simple perturbation theory.

\subsection{ General case.}

At the first glance the use of the second-case approach at arbitrary $t$ 
is more appropriate, since one can hope for better account of the 
long-range spin waves into the direct hole-hole interaction and 
for small effective formfactor, which will allow to omit hole-magnon 
vertex at all and to get 
the simplest (\ref{N5}) effective model. At the large $t/J$ ($>1$) 
it is not correct, i.e.  this transformation cannot reduce the initial 
hole-magnon vertex exactly to zero and about a half of the 
long-range spin-wave interaction remains untransformed.  Therefore, 
the rest of the hole-magnon interaction
cannot be neglected and the general form of the second-case effective 
Hamiltonian coincides with the first-case one.  This is due to the 
difference of the Green function from the simple $(-1/\omega_{\bf q})$ form 
at $t/J>1$.  

At the same time, the "first-case" approach is better in a certain sense, 
since the resulting long-range part of the hole-hole
interaction is separated 
from the short-range one, i.e. $V^{hh}_{\bf k, k^{\prime}, q}$ at ${\bf 
q}\rightarrow 0$ tends to zero and the whole long-range part
of interaction is from the 
rest of the single-magnon exchange.  We focus on the long-range part 
of interaction since, as it was found earlier \cite{SK}, it is the key 
pairing interaction for the $d_{x^2-y^2}$ two-hole bound state.
Thus, for the consideration of the two-hole problem correct account 
of the long-range magnon exchange is very important.  In the situation 
when we technically can try the energy and formfactor up to the sixth 
order of the transformation, while the hole-hole  vertex up to the fourth, 
the "first case" approach ($f_{\bf k, q}\Rightarrow f_{\bf k}$) becomes 
preferable. 

Shortly, we have performed the transformation using both approaches 
without significant differences in results, except the long-range magnon 
exchange amplitude, which 1.5 times less for the second case.  Further we 
will discuss only the "first-case" results.

Thus, we transform the initial Hamiltonian $H=H_{t-J}$ to an effective one 
$H_{eff}$
\begin{eqnarray}
\label{N6a}
&&H_{eff}=e^{-S}H e^S =H+[H,S]+\frac{1}{2!}[[H,S]S]+\dots,
\nonumber \\
&& \mbox{with}\\
&&S=  
\sum_{{\bf k},{\bf q}}f_{\bf k} M_{{\bf k},{\bf q}}\left(h_{\bf 
k-q}^{\dag} h_{\bf k} a_{\bf q}^{\dag} - \mbox{H.c.}\right)\ ,  \nonumber
\end{eqnarray}
which has the form of one in Eq. (\ref{6C}) with parameters presented in 
Appendix A.

For the sake of simplifying the consideration of the technical details 
of our approach let us omit for a moment the fifth and sixth order terms 
in the general formulae for the hole energy (see Appendix A).
Minimization of the average
 energy by variation over the TP $f_{\bf k}$
leads to the following integral equation:
\begin{eqnarray}
\label{N7}
\frac{\delta }{\delta f_{\bf k}}\left(
 \sum_{\bf k^{\prime}} E_{\bf k^{\prime}}\right)
 \sim &&I_{\bf 
k}+f_{\bf k}F^{\it 1}_{\bf k} - 2f_{\bf k}^2 I_{\bf 
k}^2-{\textstyle\frac{1}{6}}
F^{\it 2}_{\bf k}+{\textstyle\frac{1}{6}}f_{\bf k}^2 I_{\bf 
k}Y_{\bf k}\\
&&+ \frac{2J}{t}\left[f_{\bf k}\widetilde{I}_{\bf k} + 
{\textstyle\frac{1}{4}} f_{\bf k} F^{\it 3}_{\bf k} - 
{\textstyle\frac{2}{3}} f_{\bf k}^3 \widetilde{I}_{\bf k} I_{\bf 
k}- {\textstyle\frac{1}{12}} f_{\bf k} F^{\it 4}_{\bf k} +
{\textstyle\frac{1}{4}} f_{\bf k}^3 \widetilde{I}_{\bf k} Y_{\bf k}-
{\textstyle\frac{1}{12}} f_{\bf k}^3 I_{\bf k} \widetilde{Y}_{\bf k} 
\right] 
= 0 \ , \nonumber
\end{eqnarray}
where we use auxiliary functions $F^{\it n}_{\bf k}$ and $Y_{\bf k}$,
$\widetilde{Y}_{\bf k}$ given in Appendix A and
 notations introduced in the previous Section:
\begin{eqnarray}
\label{N8}
&&I_{\bf k}=\sum_{\bf q}M_{\bf k,q}^2=A_0 + 
A_{1,0}\gamma_{\bf k}^2+A_{1,1}(\gamma_{\bf k}^-)^2\ 
, \ \ \ \ \widetilde{I}_{\bf k}=\sum_{\bf q}M_{\bf 
k,q}^2\omega_{\bf q}= 
\widetilde{A}_0 + \widetilde{A}_{1,0}\gamma_{\bf 
k}^2+\widetilde{A}_{1,1}(\gamma_ {\bf k}^-)^2\ , 
\end{eqnarray} 
with the short-hand notations $\gamma_{\bf 
k}=(\cos(k_x)+\cos(k_y))/2$, $\gamma_{\bf 
k}^-=(\cos(k_x)-\cos(k_y))/2$ and numbers:
\begin{eqnarray}
\label{N9}
&&A_0=4.3273  \ ,  \ \ A_{1,0}=-3.5907  \ , \ \ A_{1,1}=-0.2453   \ , 
\\ &&\widetilde{A}_0=3.7828  \ ,  \ \ \widetilde{A}_{1,0}= 
-3.2808  \ , \ \ \widetilde{A}_{1,1}=-0.1496   \ .
\nonumber 
\end{eqnarray}
We use the following way of solving such a type of integral 
equation (\ref{N7}).
According to the discussion of the symmetry properties of the 
TP $f_{\bf k}$ given in Section II and using the 
form of  Eq. (\ref{N7}) it is evident that $f_{\bf k}=f_{\bf 
-k}=f_{{\bf k}+(\pi,\pi)}=f(k_x \leftrightarrow k_y)$, and 
hence, it can be expressed as a power series in 
$\cos(k_x)^2$, $\cos(k_y)^2$, and $\cos(k_x)\cos(k_y)$, or more 
convenient
\begin{eqnarray}
\label{N10}
f_{\bf k}=\sum_{n \le m}^{\infty} C_{n,m} \gamma_{\bf 
k}^{2(n-m)}(\gamma_{\bf k}^-)^{2m}=C_{0,0}+C_{1,0}\gamma_{\bf 
k}^2+C_{1,1}(\gamma_{\bf k}^-)^2+\dots
\end{eqnarray}
then, substituting this form of $f_{\bf k}$ in expressions for 
the auxiliary functions $F^{\it n}_{\bf k}$
 (Appendix A), one gets  an infinite number of integrals of 
the type $\sum_{\bf q}\left(M_{\bf k, q}^2 \gamma_{\bf 
k-q}^{2(n-m)}(\gamma_{\bf k-q}^-)^{2m}\right)$, each of 
them is 
a finite series in $\gamma_{\bf k}^2$ $(\gamma_{\bf k-q}^-)^2$ 
of the power $(n+2)$. Cutting $f_{\bf k}$ and all other series 
at the finite power $n$ one gets from Eq. (\ref{N7}) a set of 
$(n+1)(n+2)/2$ nonlinear algebraic equations on $C_{i,j}$ 
($i\le n$). Thus, the integral equation (\ref{N7})
is transformed to the set of algebraic equations, which is much 
easier to solve.  Keeping in mind $1/z$ character of the 
energy expansion one can hope that only a few first terms are 
important, and the role of the higher orders is insignificant. 

We solved these systems of equations numerically for the
particular values of $0<t/J<5$, and found that extension of the 
series in Eqs.  (\ref{N7}), (\ref{N10}) from $n=3$ ($\cos^6$, 10 
equations) to $n=5$ ($\cos^{10}$, 21 equations) changes 
results for the parameter $f_{\bf k}$, energy and formfactor 
(coefficients in their series) for the relative value less than 
0.5\%. Note, that including of the fifth and sixth order terms 
into expression of the energy (see Appendix A) change 
results for approximately 10\%. In all further calculations we 
used the largest ($n=5$) set of equations.
 
With the solution for $f_{\bf k}$ of such a high accuracy in 
hand one can get explicit expressions for the energy, 
formfactor, hole-hole, and hole-two-magnon vertices for the 
effective Hamiltonian Eq. (\ref{6C}).
 Evidently, as in the small $t$ limit, the hole energy has the 
shape with the minima at $\pm(\pi/2,\pm\pi/2)$ points and large 
effective mass along MBZ boundary, and  obeys the symmetry property $E_{\bf 
k}=E_{{\bf k}+(\pi,\pi)}$.

The next step of our consideration is to prove the negligible role of 
the hole-two-magnon vertices and the perturbative character of the 
renormalized  hole-magnon one.
We have calculated the second-order corrections to the single-hole energy 
from the one-magnon and two-magnon self-energy diagrams for the various 
$t/J$.  Briefly, correction to the depth of the band
from the rest of the hole-magnon vertex is no 
more than 10\%, while correction from the 
hole-two-magnon vertex (\ref{6C}) is of the next order of smallness. Namely, 
for $t/J=3$ $E_{(\pi/2,\pi/2)}=-2.22t$, $\delta E^{(1)}=-0.15t$, and 
$\delta E^{(2)}=-0.02t$.  For the correction to the effective hole-hole 
vertex the relative contribution of the hole-two-magnon exchange is even 
smaller. Single magnon exchange is really negligible for the large transfer
momentum, but is very important for the small one, indeed it has a 
"quasi-singular" form $\sim t (q_x+q_y)^2/q^2$ for the holes near the bottom 
of the band. As it was discussed 
above, this separation of the contributions to the effective 
hole-hole interaction in the momentum space leads from the construction of 
the transformation.  Note also, that the two-magnon exchange cannot provide 
the singular interaction anywhere.

Therefore, our analysis has shown the negligible role of the higher 
magnon vertices and proved that the using of the rather general type 
of the transformation leads to the transfer of the initially strong 
hole-magnon interaction mainly into the single-hole dispersion and hole-hole 
interaction. Thus, for a wide region of $t/J$ and with the high level of 
accuracy one can restrict oneself by consideration of the effective 
Hamiltonian (\ref{8C})
\begin{eqnarray}
\label{N11}
H_{eff}=\sum_{\bf k}E_{\bf k}
h_{\bf k}^{\dag} h_{\bf k}+ 2J\sum_{\bf q} \omega_{\bf q}
a_{\bf q}^{\dag} a_{\bf q} 
+ t \sum_{{\bf k},{\bf q}}F_{\bf {\bf k},{\bf q}} 
M_{{\bf k},{\bf q}}\left(h_{\bf k-q}^{\dag} h_{\bf k} 
a_{\bf q}^{\dag} + \mbox{H.c.}\right) 
+ t \sum_{{\bf k},{\bf k^{\prime}},{\bf q}} 
V^{hh}_{{\bf k},{\bf k^{\prime}},{\bf q}} 
h_{\bf k-q}^{\dag} h_{\bf k^{\prime}+q}^{\dag} 
h_{\bf k^{\prime}} h_{\bf k} \ ,
\end{eqnarray}
with all quantities defined in the Appendix A, expressed through $\mu_{\bf 
k,q}=f_{\bf k}M_{\bf k,q}$, where $f_{\bf k}$ is defined from the integral 
equation of the type of Eq. (\ref{N7}).

Figures \ref{fig3}-\ref{fig5} represent our results for the 
bottom and width of the single-hole band, and for the weights of the 
components of the magnetic polaron wave function together with ones 
from the SCBA calculations from Refs. \cite{horsch}, \cite{rhorsch}. From 
these figures we notice that the bottoms slightly differ for the all 
region of $t/J$, whereas the bandwidth difference is absent at small $t$
and tends to rise at larger $(t/J)>5$. 
 The gap between the curve corresponding to our result and SCBA 
one in Fig.  \ref{fig3} is obviously due to the absence of the long-range 
magnon contribution in the former.  Taking into account the second-order 
correction to the energy from the long-range magnon virtual 
emission-absorption provides the depth of the band $E_{(\pi/2,\pi/2)}$ and 
difference $E_{(\pi,0)}-E_{(\pi/2,\pi/2)}$, which almost coincide with the 
SCBA ones.  Surprisingly, at small $t/J$ ($<1$, $>0.5$) results for 
bandwidth are very close to each other (Fig.  \ref{fig4}).
        
Figure \ref{fig6} shows $t/J$ dependence of the hole-magnon formfactor 
$F_{\bf k,q}$ at different momenta ${\bf k}$ and ${\bf q}$. Following the 
general procedure of expansion in a series one can get the general form of 
the formfactor 
\begin{eqnarray} 
\label{N12} 
F_{\bf 
k,q}=&&\sum_{n,m,i,j}^{\infty}\left[ \left(W_{n,m,i,j}+ \omega_{\bf 
 q}W^{\omega}_{n,m,i,j}\right) \gamma_{\bf k}^{2(n-m)}(\gamma_{\bf 
k}^-)^{2m}\gamma_{\bf k-q}^{2(i-j)}(\gamma_{\bf k-q}^-)^{2j}\right]\\ 
&&=W_{0,0,0,0}+W_{1,0,0,0}\gamma_{\bf k}^2
+W_{0,0,1,0}\gamma_{\bf k-q}^2+\dots
+\omega_{\bf q}\left(W^{\omega}_{0,0,0,0}+
W^{\omega}_{1,0,0,0}\gamma_{\bf k}^2+W^{\omega}_{0,0,1,0}
\gamma_{\bf k-q}^2 +\dots\right) \ . \nonumber
\end{eqnarray}
Thus, our formfactor has strong ${\bf k}$ and ${\bf q}$-dependence at 
any $t/J$ that differs from the result of Ref. \cite{svert} where the some 
kind of variational approach has been used.  Namely, at $t/J=3$ $F_{\bf 
k,q}$ changes from 0.4 at small ${\bf q}$ to 0.0 at large 
($\sim\pi/2,\pi/2$) one.

\section{Two-hole problem.}

\subsection{Two-sublattice representation.}

Because of the AF long-range order there are two types of fermion and boson 
excitations associated with two sublattices. For considering the 
one-particle subspace it is of no importance whether one has the model with 
two different sorts of quasiparticles having the same properties 
(energy, wave function etc.), or the model with one type of quasiparticles. 
In the above consideration we used the latter for the sake of simplifying 
the notations.  One can easily prove the formal equivalence of these 
approaches.  For the two-sublattice representation there are two types of 
holes and magnons both defined in the first magnetic Brillouin zone, whereas 
for the one-sublattice representations holes and magnons are defined inside 
the full Brillouin zone, because of that $E_{\bf k}=E_{{\bf k}+(\pi,\pi)}$. 

For the calculation of the 
correlation function \cite{rhorsch}, consideration of the hole-hole 
interaction \cite{BCDS}, or some other calculations into
 two-hole subspace one should 
turn back to the two-sublattice representation. It is convenient to do 
it using the following expressions of the $h_{\bf 
k}$ and $a_{\bf q}$ operators through 
the linear combinations of the new operators
\begin{eqnarray}
\label{T1}
&&h_{\bf k}=(f_{\bf k}+g_{\bf k})/\sqrt{2}\ , \ \ \
h_{{\bf k}+(\pi,\pi)}=(f_{\bf k}-g_{\bf k})/\sqrt{2}  \ , \\
&&a_{\bf q}=(\alpha_{\bf q}+\beta_{\bf q})/\sqrt{2}\ , \ \ \
a_{{\bf q}+(\pi,\pi)}=(\alpha_{\bf q}-\beta_{\bf q})/\sqrt{2} \ ,
\nonumber
\end{eqnarray}
where $f_{\bf k}$ and $g_{\bf k}$ correspond to the fermionic excitations at
the $A$ and $B$ sublattice, respectively. $\alpha_{\bf q}$ and $\beta_{\bf 
q}$ are the two type of Bogolubov spin-wave excitations.
Transition to these new variables for the hole-magnon 
part of the effective Hamiltonian is straightforward 
\begin{eqnarray}
\label{T2}
H^{ha}_{eff}\Rightarrow t \sum_{{\bf k},{\bf q}}F_{\bf {\bf k},{\bf q}} 
M_{{\bf k},{\bf q}}\left(f_{\bf k-q}^{\dag} g_{\bf k} 
\beta_{\bf q}^{\dag} + g_{\bf k-q}^{\dag} f_{\bf k} 
\alpha_{\bf q}^{\dag} + \mbox{H.c.}\right) \ ,
\end{eqnarray}
where summation is performed over the MBZ.

Now we should obtain $h, h \rightarrow f, g$ transformation of the 
hole-hole interaction part of the effective Hamiltonian (\ref{N11}).  
Rewriting the $hh$ interaction term (\ref{N11}) in the form where
summation is produced over MBZ leads to
\begin{eqnarray}
\label{T3}
H^{hh}_{eff}\Rightarrow H^{fg}+H^{ff}+H^{gg}=
 t \sum_{{\bf k},{\bf k^{\prime}},{\bf q}}
V^{fg}_{{\bf k},{\bf 
k^{\prime}},{\bf q}} f_{\bf k-q}^{\dag} 
g_{\bf k^{\prime}+q}^{\dag} g_{\bf k^{\prime}} f_{\bf k} +
 t \sum_{{\bf k},{\bf k^{\prime}},{\bf q}}\left( V^{ff}_{{\bf k},{\bf 
k^{\prime}},{\bf q}} f_{\bf k-q}^{\dag} f_{\bf k^{\prime}+q}^{\dag} 
f_{\bf k^{\prime}} f_{\bf k} 
+ (f\rightarrow g)\right)\ .
\end{eqnarray}
Thus, there are three different parts in the $H_{eff}^{hh}$, 
which correspond to the interaction between holes at the different 
sublattices ($fg$-part) and at the same one ($ff$- and $gg$-parts). The 
first terms for the interaction of the particles at the same sublattice 
arise in the third order of the transformation and physically correspond to 
the process shown in the diagram Fig. \ref{fig7}a. At the Ising background 
it is equivalent to the four-hopping process of the two holes staying at the 
apart corners of the minimal lattice square, when they move in the clockwise 
or anticlockwise direction doing the half of a turn. Generally, $ff$ or 
$gg$ interaction has no some important features of the $fg$ one. Namely, 
there are no singularities in their long-range interaction as well as 
the gain of the energy due to reducing of the number of broken AF bonds 
is absent.  These physical reasons were checked earlier \cite{CDS} and it 
was found that the bound states are absent for the particles at the same 
sublattice in the region of $(t/J)>1$.  Thus, we will restrict ourselves by 
consideration of the interaction of the particles at the different 
sublattices. 

For obtaining $fg$ interaction from $hh$ one accurate 
consideration of the parity of the vertex $V^{hh}$ under the transformation
$R= {\bf k}({\bf k^{\prime}})\rightarrow {\bf k}({\bf k^{\prime}})+
(\pi,\pi)$ is required. 
There are two contributions of different parity ($R=\mp$) in the effective 
$fg$ interaction.  Their diagrammatic analogues are presented in Figs.  
\ref{fig7}b and \ref{fig7}c, respectively.  The first one is due to the
one-magnon exchange and by its origin it is of the "exchange" type 
($V^{hh}_{ex}$). The second one is due to the two-magnon exchange and the 
contact interaction (additions to $J$-term) and it is of "direct" type 
($V^{hh}_{dir}$).  Obviously, these contributions enter in the $fg$ vertex 
with different sign 
\begin{eqnarray} 
\label{T4} 
V^{fg}_{{\bf k},{\bf 
k^{\prime}},{\bf q}}= \left(-V^{hh}_{ex}({\bf k},{\bf k^{\prime}},{\bf 
q})-V^{hh}_{ex}({\bf k^{\prime}},{\bf k},{\bf -q})+ V^{hh}_{dir}({\bf 
k},{\bf k^{\prime}},{\bf q}) + V^{hh}_{dir}({\bf k^{\prime}},{\bf k},{\bf 
-q}) \right)\ .  
\end{eqnarray} 
Note here, that the first non zero 
correction beyond the ladder approximation for the hole-hole ($fg$) 
scattering arises only in the sixth order over $t$ (see Fig. \ref{fig7}d), 
that is the same order as one of the Trugman processes for the single hole 
movement and, as it seems, it has the similar reasons in the basis. 
Therefore, keeping in mind the negligible role of these non SCBA 
contributions to the hole energy, one can hope that the role of the diagram 
in Fig.  \ref{fig7}d in the hole-hole interaction can be omitted and the 
ladder approximation will work well even for the initial 
(untransformed) $t$-$J$ model.  In our calculations we use the same 
approximation, but previously obtain the effective vertices, which are 
perturbative. 

\subsection{Types of pairing interaction.}

Generally, there are two different types of the hole-hole interaction in the 
$t$-$J$ model. The first one is due to the spin-wave exchange, whereas the 
second is from minimization of the number of broken AF bonds by the placing 
of the holes at the nearest neighbor sites. We will consider them 
separately.

The second type of interaction is
usually introduced in the pure $t$-$J$ model by 
adding projectors $P_i=(1-n^h_i)$ in $J$-term
\begin{eqnarray}
\label{T5}
H_{J}= J \sum_{i,j}\left[(1-n^h_i) {\bf S}_i{\bf S}_j 
(1-n^h_j) - {\textstyle\frac{1}{4}}n^h_i n^h_j \right]\ , 
\end{eqnarray}
which  project out the subspace of local states (spins),
 $n^h_i = h_i^{\dag}h_i$ is 
the operator of the hole number. It is evident that due to the $h_i^{\dag} 
(h_i)$ and ${\bf S}_i$ operator commutativity projection procedure 
is exact, i.e. there is no spin-spin interaction between the sites with 
the holes. Thus, the addition part to the $t$-$J$ model can be written as
\begin{eqnarray}
\label{T6}
\Delta H_{t-J}= J \sum_{i,j}\left(-(n^h_i+n^h_j){\bf S}_i{\bf S}_j+
n^h_i n^h_j{\bf S}_i{\bf S}_j - {\textstyle\frac{1}{4}}n^h_i n^h_j\right)\ ,
\end{eqnarray}
where summation run over bonds. Further treating this interaction term 
in the spin-wave approximation is straightforward and yields 
\begin{eqnarray} 
\label{T7}
\Delta H_{t-J}= -2J(1-2\delta\lambda) 
\sum_{{\bf k},{\bf k^{\prime}},{\bf q}}\gamma_{\bf q}
 f_{\bf k-q}^{\dag} g_{\bf k^{\prime}+q}^{\dag} g_{\bf k^{\prime}} 
f_{\bf k} + \delta H_{J} \ ,
\end{eqnarray}
where term $\delta H_{J}$, consisting of the two-magnon terms $n^h aa$ 
and $n^h n^h a a$, is presented in Appendix B. Hole attraction 
is enhanced by the zero-point fluctuations by the constant 
$(-2\delta\lambda\simeq 0.16$). Applying the transformation (\ref{N6a}) to 
the Hamiltonian  (\ref{T7}) one can get the addition part of the 
effective Hamiltonian
\begin{eqnarray}
\label{T7a}
\delta H^{fg}=
 J \sum_{{\bf k},{\bf k^{\prime}},{\bf q}}\delta V^{fg}_{{\bf k},{\bf 
k^{\prime}},{\bf q}} f_{\bf k-q}^{\dag} g_{\bf k^{\prime}+q}^{\dag} 
g_{\bf k^{\prime}} f_{\bf k} \ .
\end{eqnarray} 
The expression of the second-order (in $f_{\bf k}$) addition to the 
bare vertex (\ref{T7}) is cumbersome and we present the general expression 
of $\delta V^{fg}_{{\bf k},{\bf k^{\prime}},{\bf q}}$ in Appendix B.

Generally, there is an evident result for the $n_i n_j$ interaction in the 
$t=0$ limit, namely, the groundstate of two holes is the bound state with the 
energy $E_b\simeq -0.58J$. Moreover, there is the full degeneracy among 
the states of $s$ [$\cos k_x+\cos k_y$], $d$ [$\cos k_x-\cos k_y$],   and 
$p$ [$\sin k_x,$ $\sin k_y$] symmetries. 
The $n n$ type of interaction (\ref{T7}) was intensively studied by the 
number of analytical  \cite{Eder3}, \cite{CDS}, and numerical techniques 
\cite{Poil1},  which established that increase of $t$ (taking 
into account the spin-wave exchange interaction) leads to the rapid 
growth of $E_b$ and disappearance of bound states.  It was found that  
the largest critical value of $t_c$ is for the $d$ symmetry of the bound 
state and that it is somewhere between $t=(2-5)J$, which is very close to 
the values of $t$ proposed for the real CuO$_2$ planes. This fact has
stimulated the developing of the idea of the superconductivity by 
the "preformed" pairs (bosons). 

Considered pairing interaction has nothing to do with the spin-fluctuation 
one, which was investigated in Refs. \cite{Pines1} on the 
phenomenological basis and in Refs. \cite{Schr1}  using the 
RPA for the Hubbard model. An essential addition to the studying of the 
spin-wave exchange (Fig. \ref{fig7}b) in the $t$-$J$ and Hubbard models has 
been done in the work by Frenkel and Hanke \cite{Fren}, where authors found 
that the exchange by the long-range (small momentum transfer) spin wave 
leads to the dipolar interaction between holes, which can be attractive or 
repulsive depending on the relative location of the holes. In the later work 
by Kuchiev and Sushkov \cite{SK} this problem was independently studied in 
great details and several interesting features of the system were found. 
First of all, neglecting the retardation effect and finite size of the 
Brillouin zone for the two-hole problem one can obtain the Schr\"{o}dinger 
equation with the potential $\sim (x^2-y^2)/r^4$, which leads to the "fall 
to the center" effect and to the infinite number of bound states.  It was 
also found that the actual deepness of the bound states is very sensitive to 
the curvature of the hole band along the MBZ boundary and that for the equal 
masses in directions along and perpendicular to MBZ boundary there are no 
bound states at all. This effect is explained by the effective lowering of 
the dimensionality of the system when the mass along MBZ boundary tends to 
infinity, that is led to the strengthening of the pairing interaction.  In 
Ref. \cite{SK} only $d$ and $g$ [$(\cos k_x-\cos k_y)\sin k_x\sin k_y$] 
states were found to exist, and the lowest one was the $d$-state.
This confirms a general statement \cite{Hirsch} that in AF state 
one-magnon exchange leads to the repulsion in $s$-wave and attraction 
in $d$-wave state.

Note, that due to the long-range nature of the interaction and the large 
size of the proposed bound states it is not obvious how cluster 
analysis can directly reveal them.

\subsection{Bethe-Salpeter equation.}

Thus, one obtains the Hamiltonian (\ref{N11}) with the hole-magnon 
(\ref{T2}) and "contact" hole-hole (\ref{T4}), (\ref{T7a}) interactions.  
Since we turned to an effective Hamiltonian by canonical transformation 
(\ref{N6a}) the short-range spin-wave exchange (Fig. \ref{fig7}b) is 
now included in the "contact" interaction, which does not 
contain the retardation.
Both interactions, according to above discussion, can lead to the pairing. 
As it was noted, the correct account for the retardation effect in the 
spin-wave exchange diagram is important, so let us consider this problem 
in details.

The systematical method to search the bound states 
is to look for the two-particle Green function poles in the 
scattering channel as the function of the 
particles summary energy in the system 
of their center of inertia \cite{LL4}.  
Corresponding integral equation for two holes with the total momentum ${\bf 
P}=0$ is presented in Fig. \ref{fig8a} in a diagrammatic form.
The standard way of solving of such a type of equation with the non-retarded 
"compact" bare vertex $\Gamma^0$ is given in Appendix C.

In our case the "compact" vertex
$\Gamma^0$ consists of two parts (see Fig. \ref{fig8b}) and one 
has to include the magnon propagator into expression for 
the spin-wave exchange vertex.
A natural assuming that the two-particle Green function has no singularities 
as the function of the difference of the energies of incoming particles 
provides somewhat different way of solving the Bethe-Salpeter problem. 
Details are also given in Appendix C. 

Resulting parametric equation on the bound state energy 
$E$ of the Bethe-Salpeter type for the problem with two vertices (Fig.  
\ref{fig8b}), which we will solve looks like
\begin{eqnarray}
\label{T17}
\psi({\bf k},E)=\frac{1}{E-2E_{\bf k}} \sum_{\bf p} \left[
\frac{-2(F_{\bf k, k+p} M_{\bf k, k+p})^2
}{E-E_{\bf p}-E_{\bf k}-\omega_{\bf q}}+V^{fg}_{{\bf k},-{\bf k},{\bf q}}
+ \delta V^{fg}_{{\bf k},-{\bf k},{\bf q}}\right]
\psi({\bf p},E) \ ,
\end{eqnarray}
where ${\bf q}={\bf k}\pm{\bf q}$ for the exchange (direct) parts of 
$V^{fg}$ (\ref{T3}) and $\delta V^{fg}$ (\ref{T7a}).

\subsection{Results.}

Thus, having in hands the vertices (\ref{T2}), (\ref{T4}), (\ref{T7a}) and 
equation (\ref{T17}) one can hope to obtain  reliable results for the bound 
states in the $t$-$J$ model.  Moreover, since we have done the separation of 
the scales in the momentum space and consider the "minimization of the 
number of broken bonds" interaction separately, one can demonstrate the role 
of each type of interaction in forming of the bound states. 

Shortly, our results are as follows. The single bound state of $d$ symmetry
 ($d_{x^2-y^2}$) is exist in all region of $0<(t/J)<5$. The states 
of other symmetries ($s$, $p$) at $(t/J)\ge 0.2$
were not found. The main thesis of this 
work is that the interplay of both interactions, which tend to $d$-wave 
pairing, namely the short-range $J$-term (\ref{T7a}) and the long-range 
spin-wave exchange (\ref{T2}), is important for the formation of this state. 
 Specifically, there are no bound states from the $J$-term 
alone for $(t/J)>2.1$. The spin-wave exchange itself provide the 
shallow bound state, but the actual magnitude of the mass along 
the MBZ boundary almost pushes out this state in the continuous 
spectra.  Nevertheless, putting these interactions together we 
get the bound state with the energy of the two orders of 
magnitude deeper than one found in Ref.  \cite{SK}. 
It was claimed in Ref. \cite{SK} that due to the $1/r^2$ nature of 
the attractive potential accurate account for the "cutoff" 
parameter (finite size of the MBZ) and the retardation effect is required.  
We have taken into account both effects and have done calculations for the
lowest bound state.
Deviation 
of the model from the pure $t$-$J$ one, i.e. taking into account 
next hopping integrals, can lead to the different physics of the 
system.

Let us discuss the other symmetry states. As it was noted above at 
the limit $t=0$ the $J$-term bound states of $d$, $s$, and $p$ symmetry are 
not distinguished in energy, but already at $(t/J)=0.1$ $s$- and $p$-states
are two times higher than the $d$-state and they are disappeared at 
$(t/J)=0.2$. 

Considering terms in equation in the bound state energy (\ref{T17}) 
separately and all together, we have obtained results for the $d$-wave 
pairing state shown in Fig. \ref{fig9}. The dashed curve corresponds to 
taking into account the transformed $J$-term of interaction (\ref{T7a}) 
alone ($\delta V^{fg}_{{\bf k},-{\bf k},{\bf q}}$).  Obtained critical value 
of $t$ for disappearing this, short-range in nature, state $t_c=2.1J$ is in 
excellent agreement with the  variational approach 
\cite{Eder3}, and in the good one with the
finite-cluster calculations \cite{Dag}, \cite{Poil1}, and 
other approaches \cite{CDS}. Dashed-dotted curve corresponds to the 
long-range bound state due to the first two terms in Eq. (\ref{T17}).  
According to Ref. \cite{SK} this state should have small 
negative energy.  We found that if one takes into account the 
second-order correction to the single-hole energy due to the 
rest of the hole-magnon vertex (\ref{N11}) it will lead to the 
slightly smaller mass along MBZ boundary and then to the very small
binding energy $\sim -(10^{-3}-10^{-4})t$ for the long-range 
state.  Actually, corrected value of the mass almost pushed up 
the bound state in the continuous spectra.  As we noted in Sec.  
IV resulting masses coincide with the SCBA ones.  The solid 
curve is our final result for the $d$-wave bound state energy in 
the $t$-$J$ model.  Bound state energy for $(t/J)=3$ equals to 
$\Delta E=E-2E_{\bf k_0}=-0.022t$, which is of the two 
orders of magnitude deeper than it was obtained earlier \cite{SK}.  Thus we 
obtained the strong enhancement  of the coupling effect because of the 
interplay of both types of pairing interaction.  

Note, that the "contact" part of the 
spin-wave exchange interaction ($V^{fg}$) plays the minor role in such a 
strong effect, namely ignoring it in Eq. (\ref{T17}) yields the energy 
$-0.01t$, which is only two times smaller than the result of the integral 
effect.

It is useful to consider the structure of the wave functions of the two 
hole bound states in the ${\bf k}$-space. Figure \ref{fig10}a shows the 
wave function for $(t/J)=1$. It is simply "bare" short-range
$\psi_{\bf k} \sim (\cos{k_x}-\cos{k_y})$ with the small addition of the 
higher harmonics.  Figures \ref{fig10}b,c,d  show wave functions for the (b) 
long-range state, (c) short-range one, and (d) resulting wave 
function, all for $(t/J)=2$.  Whereas the long-range bound state 
\ref{fig10}b is well localized near the band minima, the 
short-range one \ref{fig10}c looks like something average between Figs. 
\ref{fig10}a and \ref{fig10}b. It is not easily understandable, 
but one can say that in canonically transformed vertex $\delta 
V^{fg}$ (\ref{T7a}) except the simple $\gamma_{\bf q}$ term 
there are some second-order terms with the same kinematic 
structure as in the $V^{fg}$ vertex, and since at $(t/J)=2$ 
this state is shallow the higher harmonics with this structure 
become important.  Resulting wave function \ref{fig10}d reveals 
the features of the previous states.  We believe that our 
results are very close to exact ones in all region of primary 
interest $1<(t/J)<5$.  Lowering of the bound state energy for 
the $(t/J)=5$ manifests the worse accuracy of our approach for 
the very large $t$.

The next problem, which can be touched, is the influence of the next-nearest 
hopping terms ($t^{\prime}$-terms) on the bound states. Evidently,   
small $t^{\prime}$ leads to the perturbative addition to the 
hole dispersion $\delta E_{\bf k}=4t^{\prime}\cos(k_x)\cos(k_y)$ 
and for the positive value of $t^{\prime}$ it makes the band 
more flat in ($\pi/2,\pi/2$)$\rightarrow$($\pi,0$) direction.
  According to the above discussion it also strongly 
enhances the long-range interaction and makes the $d$-wave bound 
state much deeper. For instance, for the absolutely flat band 
($m_{\|}=\infty$) at $(t/J)=3$ energy of the bound state is 
$E=-0.165t$. Note, that the neglecting of the short-range 
interaction ($J$-term) provides result of the next order of 
 smallness $E=-0.023t$. Further increase of $t^{\prime}$ will 
 lead to the shift of the minima to the $(0,\pm\pi)$, 
$(\pm\pi,0)$ points and will make the bound state shallower. For 
some region of $t^{\prime}>0$ the long-range bound state of $g$ 
symmetry becomes possible.  Its wave function obeys the same 
symmetry as [$(\cos k_x-\cos k_y)\sin k_x\sin k_y$], i.e changes 
the sign in MBZ eight times. Since there is no short-range 
interaction for this state, the energy associated with it is 
very small.

It is well established by now, that for the real CuO$_2$ compounds 
$t^{\prime}$ has the negative sign and the total effect from  
$t^{\prime}$-terms is the fully isotropic dispersion near the band minima 
\cite{Bala,BCS}. Note, that it is the main effect from such terms 
and one can neglect their contribution to the effective vertices 
up to the rather large values of $t^{\prime}$. Joining this 
statement together with the claimed sensitivity of the bound 
state to the anisotropy of the band one can suggest that there 
are no bound states in the realistic model of the CuO$_2$ plane. 
We have studied the problem of the critical value of 
$t^{\prime}$ and found that the difference 
$\Delta=E_{(\pi,0)}-E_{(\pi/2,\pi/2)}$, which is directly connected with the 
inverse mass along MBZ boundary, at which the bound state energy tends to 
zero is equal to $0.22t$ for $(t/J)=3$. Comparing it with the bare $t$-$J$ 
model value of $\Delta=0.13t$ one can get the critical $t^{\prime}_c\simeq 
0.3 J$, which is much lower than the realistic value $t^{\prime}_{eff}\sim 
1.5 J$. 

Turning back to the simple $t$-$J$ model, one can say that, at the 
first glance, obtained result strongly supports the idea of the 
$t$-$J$ superconductivity by the condensate of the "preformed" 
pairs.  In our opinion such a picture is not so obvious, since 
the existence of the long-range spin wave excitations is 
essential for the bound state formation, whereas the long-range 
order is unstable under very small \cite{SS1,SF} 
finite concentration of holes. Therefore, for the clearing this 
subject one has to solve the pairing and stable spin state 
problem selfconsistently.

\section{Conclusion}

We conclude by summarizing our results. 

We have put forward the canonical transformation of the $t$-$J$ model 
Hamiltonian using an analogy with the lattice polaron problem as well as 
some evident ideas based on the known single-hole properties in the AF 
background.  We have shown that the rather wide type of the 
transformations, which has the some type of the $1/z$ expansion in the 
basis, allow to extend the region of the analytical treatment of the 
problems up to $t/J\sim 5$ with an appropriate accuracy. Generally, powerful 
method applied provided us the straight way to the formulating of the 
quasiparticle Hamiltonian, which includes the free energy terms and all 
essential interactions. 
 
The proof of the correctness of the simple form of 
the effective quasiparticle Hamiltonian has been done. Results for the 
single-hole bottom of the band, bandwidth, and other properties have been 
compared with ones of the SCBA calculations and a remarkable agreement has 
been found.  It is supported the idea that the "canonically transformed" 
quasiparticles have the properties, which are close to ones of exact $t$-$J$ 
model quasiparticles. Although the bandwidth and 
components of magnetic polaron wave function have discrepancies with the 
exact ones, which are grown for the larger $(t/J)>5$, the bottom of the hole 
band remains very close to an exact one, confirming the basic idea that the 
effect of the polaron well formation (of the order of $t$) is taken into 
account by our transformation.

Using the obtained Hamiltonian we have performed the study of the two-hole 
problem. The hole-hole interactions of the different nature have been 
considered separately and all together. Rather deep bound state of $d$-wave 
symmetry originated from the interplay of two types of the pairing 
interactions has been found. Retardation effect for the long-range spin-wave 
exchange has been carefully taken into account. Other possible symmetries 
of the bound state wave function have been studied as well. The main effect 
of the so called $t^{\prime}$-terms has been investigated and the critical 
value of $t^{\prime}_c$ for the bound state existence has been found.

Since we have used the presence of the AF long-range order as a foundation 
of the setting up the problem and the long-range interaction was found to be 
essential for the bound state formation, so the direct relation of the 
considered two-hole problem to the case of finite hole doping of the real 
CuO$_2$ plane is unclear. We have briefly discussed the 
questions, which remain to be resolved.

\vskip 1.cm
{\Large \bf Acknowledgments} \\ \vskip 0.1cm
We are grateful to E. G. Batyev for useful discussions.  
This work was supported in part by the  Russian Foundation for Fundamental 
Researches, Grant No 94-02-03235. 
 The work of one of us 
(A.L.C.) has been supported by a fellowship of INTAS Grant 93-2492 
and is carried out within the research program of International 
Center for Fundamental Physics in Moscow.
V. S. acknowledges the  support by ISSEP, Grant No. s96-1436.

\appendix
\section{}
The series of commutators of the type (\ref{N6a}) with generator of the 
transformation (\ref{4C}) provide the general expression for the hole energy 
and hole-magnon formfactor (\ref{6C})
\begin{eqnarray}
\label{AA1}
E_{\bf k}= &&2t\sum_{\bf q}\left( M_{\bf k, q}\mu_{\bf k, q}+ 
{\textstyle\frac{1}{3!}} W_{\bf k,q}\mu_{\bf k, q} + 
{\textstyle\frac{1}{5!}} \delta W_{\bf k,q}\mu_{\bf k, q} + 
\frac{2J}{t}\left[ {\textstyle\frac{1}{2!}} \mu_{\bf k, q}^2 \omega_{\bf q} 
+ {\textstyle\frac{1}{4!}} \widetilde{W}_{\bf k,q}\mu_{\bf k, q} 
+ {\textstyle\frac{1}{6!}} \delta
\widetilde{W}_{\bf k,q}\mu_{\bf k, q}\right] \ ,
\right) \\
F_{\bf k,q}= &&1+\left({\textstyle\frac{1}{2!}} 
W_{\bf k,q} + {\textstyle\frac{1}{4!}} \delta W_{\bf k,q} + 
\frac{2J}{t}\left[ \mu_{\bf k, q} \omega_{\bf q}
+ {\textstyle\frac{1}{3!}} \widetilde{W}_{\bf k,q}
+ {\textstyle\frac{1}{5!}} 
\delta \widetilde{W}_{\bf k,q}\right] \right)/M_{\bf k, q} \ , 
\nonumber
\end{eqnarray}
with
\begin{eqnarray}
\label{AA2}
W_{\bf k,q}=&3\mu_{\bf k, q}\biggl(&V_{\bf k-q}-V_{\bf k}\biggr)-
M_{\bf k, q}\biggl(U_{\bf k-q}+U_{\bf k}\biggr)\ , \ \ \ 
\widetilde{W}_{\bf k,q}=\mu_{\bf k, q}\biggl[3\biggl(
\widetilde{V}_{\bf k-q}-\widetilde{V}_{\bf k}\biggr)- 
\omega_{\bf q}\biggl(U_{\bf k-q}+U_{\bf k}\biggr)\biggr]\nonumber\\ 
\delta W_{\bf k,q}=&\mu_{\bf k, q}\biggl(15(&U_{\bf k}V_{\bf k}-
U_{\bf k-q}V_{\bf k-q})+5(U_{\bf k-q}V_{\bf k}-U_{\bf k}V_{\bf k-q})\biggr)
+M_{\bf k, q}\biggl(U_{\bf k-q}^2+U_{\bf k}^2+6 U_{\bf k-q}U_{\bf k}\biggr)\ 
,
\nonumber\\ 
&+\sum_{\bf q^{\prime}}\biggl[&
\mu_{\bf k, q}\biggl\{ 5(U_{\bf k-q^{\prime}}
M_{\bf k,q^{\prime}}\mu_{\bf k,q^{\prime}} -
U_{\bf k-q-q^{\prime}}M_{\bf k-q,q^{\prime}}\mu_{\bf k-q,q^{\prime}})
+8(V_{\bf k-q-q^{\prime}}\mu_{\bf k-q,q^{\prime}}^2-
V_{\bf k-q^{\prime}}\mu_{\bf k,q^{\prime}}^2)\biggr\}
\nonumber\\ 
&&+M_{\bf k, q}\biggl\{U_{\bf k-q^{\prime}}\mu_{\bf k,q^{\prime}}^2 
+U_{\bf k-q-q^{\prime}}\mu_{\bf k-q,q^{\prime}}^2\biggr\}
\biggr]+\delta W^{Tr}_{\bf k,q}\ , \\
\delta\widetilde{W}_{\bf k,q}=&\mu_{\bf k, q}\biggl(15(&U_{\bf k}
U_{\bf k}^{\omega}-U_{\bf k-q}U_{\bf k-q}^{\omega})
+5(U_{\bf k-q}U_{\bf k}^{\omega}-U_{\bf k}U_{\bf k-q}^{\omega})
+\omega_{\bf q}\bigl(U_{\bf k-q}^2+U_{\bf k}^2+6 U_{\bf k-q}U_{\bf k}\bigr)
\nonumber\\ 
&+\sum_{\bf q^{\prime}}\biggl[&
5(U_{\bf k-q^{\prime}}\mu_{\bf k,q^{\prime}}^2 \omega_{\bf q^{\prime}}-
U_{\bf k-q-q^{\prime}}\mu_{\bf k-q,q^{\prime}}^2\omega_{\bf q^{\prime}})
+8(U_{\bf k-q-q^{\prime}}^{\omega}\mu_{\bf k-q,q^{\prime}}^2-
U_{\bf k-q^{\prime}}^{\omega}\mu_{\bf k,q^{\prime}}^2)
\nonumber\\ 
&&+\omega_{\bf q}\bigl\{U_{\bf k-q^{\prime}}\mu_{\bf k,q^{\prime}}^2 
+U_{\bf k-q-q^{\prime}}\mu_{\bf k-q,q^{\prime}}^2\bigr\}
\biggr]\biggr)+\delta\widetilde{W}^{Tr}_{\bf k,q}\ , \nonumber
\end{eqnarray}
where
\begin{eqnarray}
\label{AA3}
&&V_{\bf k}=\sum_{\bf q'}M_{\bf k, q'}\mu_{\bf k, q'}, \ \ \ 
U_{\bf k}=\sum_{\bf q'}\mu_{\bf k, q'}^2, \ \ \ 
\widetilde{V}_{\bf k}=\sum_{\bf q'}\mu_{\bf k, q'}^2\omega_{\bf q'} \ ,
\end{eqnarray}
and $\delta W^{Tr}_{\bf k,q}$, $\delta\widetilde{W}^{Tr}_{\bf k,q}$ are the 
terms from the Trugman processes. They were found to play 
the negligible role for the energy and vertex. 

Explicit expression of the energy up to the fourth order over
the TP for the generator in the form (\ref{N6a}) is 
\begin{eqnarray}
\label{AA4}
E_{\bf k}\simeq 
&& 2t \left( f_{\bf k}I_{\bf k}+{\textstyle\frac{1}{6}}\left[
3f_{\bf k}^2 F^{\it 1}_{\bf k}
-4f_{\bf k}^3 I_{\bf k}^2-f_{\bf k}F^{\it 2}_{\bf k}\right]
+\frac{J}{t}\left[f_{\bf k}^2 \widetilde{I}_{\bf 
k}+{\textstyle\frac{1}{12}}\left(3 f_{\bf k}^2  
F^{\it 3}_{\bf k}- 4 f_{\bf k}^4 \widetilde{I}_{\bf k} I_{\bf 
  k}-f_{\bf k}^2  F^{\it 4}_{\bf k}
\right)\right]\right) \ ,
\\
\mbox{with}\hskip 1.cm
&&F^{\it 1}_{\bf k}=\sum_{\bf q}\left(M_{\bf k, q}^2 I_{\bf k-q} 
f_{\bf 
k-q}\right)\ , \ \ \ \ \ 
F^{\it 2}_{\bf k}=\sum_{\bf 
q}\left(M_{\bf k, q}^2 I_{\bf k-q} f_{\bf k-q}^2\right)\ , \nonumber
\\
&&F^{\it 3}_{\bf k}=\sum_{\bf q}\left(M_{\bf k, q}^2 \widetilde{I}_{\bf 
k-q} f_{\bf k-q}^2\right)\ ,\ \ \ \ \ 
F^{\it 4}_{\bf k}=\sum_{\bf q}\left(M_{\bf k, q}^2 I_{\bf k-q} 
f_{\bf k-q}^2\omega_{\bf q}\right)\ ,
\nonumber\\
&&Y_{\bf k}=\sum_{\bf q}\left(M_{\bf k+q, q}^2 \right)\ ,\ \ \ \ \ 
\widetilde{Y}_{\bf k}=\sum_{\bf q}\left(M_{\bf k+q, q}^2
\omega_{\bf q}\right)\ ,
\nonumber
\end{eqnarray}
where $I_{\bf k}$ and $\widetilde{I}_{\bf k}$  are defined by Eq. 
(\ref{N8}).

Hole-hole vertex $V^{hh}_{{\bf k},{\bf k^{\prime}},{\bf q}} =
V^{hh}_{ex}({\bf k},{\bf k^{\prime}},{\bf q}) +V^{hh}_{dir}({\bf k},{\bf 
k^{\prime}},{\bf q})$ is
\begin{eqnarray}
\label{AA5}
V^{hh}_{ex}({\bf k},{\bf k^{\prime}},{\bf q})=&& 
\biggl(M_{{\bf k},{\bf q}}\mu_{{\bf k^{\prime}+q},{\bf q}}
+M_{{\bf k-q},{\bf -q}}\mu_{{\bf k^{\prime}},{\bf -q}}\biggr)\biggl[
1+{\textstyle\frac{1}{3!}}P_{{\bf k},{\bf k^{\prime}},{\bf q}}\biggr]
\\
&&+ \frac{2J}{t}\omega_{\bf q}
\biggl(\mu_{{\bf k},{\bf q}}\mu_{{\bf k^{\prime}+q},{\bf q}}
+\mu_{{\bf k-q},{\bf -q}}\mu_{{\bf k^{\prime}},{\bf -q}}\biggr)\biggl[
{\textstyle\frac{1}{2!}}+
{\textstyle\frac{1}{4!}}P_{{\bf k},{\bf k^{\prime}},{\bf q}}\biggr]
\nonumber \\
&&+4\biggl(\mu_{{\bf k},{\bf q}}\mu_{{\bf k^{\prime}+q},{\bf q}}
-\mu_{{\bf k-q},{\bf -q}}\mu_{{\bf k^{\prime}},{\bf -q}}\biggr)\biggl[
{\textstyle\frac{1}{3!}}(V_{\bf k-q}-V_{\bf k})+
\frac{2J}{t}{\textstyle\frac{1}{4!}}(U_{\bf k-q}^{\omega}
-U_{\bf k}^{\omega})\biggr]\ ,
\nonumber \\
\mbox{with}\hskip 1.cm
&& P_{{\bf k},{\bf k^{\prime}},{\bf q}}=-3(U_{\bf k-q}+U_{\bf k})-
(U_{\bf k^{\prime}-q}+U_{\bf k^{\prime}})\ ,
\nonumber 
\end{eqnarray}
and 
\begin{eqnarray}
\label{AA6}
V^{hh}_{dir}({\bf k},{\bf k^{\prime}},{\bf q})=&&
T^{\it 1}_{{\bf k},{\bf k^{\prime}},{\bf q}}
+T^{\it 1}_{{\bf k-q},{\bf k^{\prime}+q},{\bf -q}}
+T^{\it 2}_{{\bf k},{\bf k^{\prime}},{\bf q}}
+T^{\it 2}_{{\bf k-q},{\bf k^{\prime}+q},{\bf -q}}\ ,
 \\
T^{\it 1}_{\bf k, k^{\prime}, q}=\sum_{\bf q^{\prime}}
\mu_{\bf k^{\prime}+q, q-q^{\prime}}\biggl[&&
\mu_{\bf k^{\prime}+q^{\prime}, q^{\prime}}
(M_{\bf k-q^{\prime}, q-q^{\prime}}
\mu_{\bf k, q^{\prime}}-M_{\bf k,q^{\prime}}
\mu_{\bf k-q^{\prime}, q-q^{\prime}})
\nonumber \\
&&-\mu_{\bf k^{\prime}, -q^{\prime}} 
(M_{\bf k-q^{\prime}, q-q^{\prime}}
\mu_{\bf k-q^{\prime}, -q^{\prime}}
+M_{\bf k-q^{\prime}, -q^{\prime}}
\mu_{\bf k-q^{\prime}, q-q^{\prime}})\biggr]
\nonumber \\
T^{\it 2}_{\bf k, k^{\prime}, q}=-\sum_{\bf q^{\prime}}
\mu_{\bf k^{\prime}+q, q^{\prime}}&&
\mu_{\bf k-q+q^{\prime}, q^{\prime}}\biggl[
M_{\bf k, q-q^{\prime}}\mu_{\bf k^{\prime}+q-q^{\prime}, q-q^{\prime}}
+M_{\bf k-q+q^{\prime}, -q+q^{\prime}}
\mu_{\bf k^{\prime},-q+q^{\prime}}
\nonumber \\
&&+ M_{\bf k^{\prime}+q-q^{\prime},q-q^{\prime}}
\mu_{\bf k, q-q^{\prime}}
+M_{\bf k^{\prime},-q+q^{\prime}}
\mu_{\bf k-q+q^{\prime}, -q+q^{\prime}}
\biggr]
\nonumber 
\end{eqnarray}
Two-magnon vertices are
\begin{eqnarray}
\label{AA7}
V^{haa}_{1}({\bf k},{\bf q},{\bf q^{\prime}})=&& 
M_{{\bf k-q},{\bf q^{\prime}}}\mu_{{\bf k},{\bf q}}
 -M_{{\bf k},{\bf q}}\mu_{{\bf k-q},{\bf q^{\prime}}} +
\frac{J}{t}(\omega_{\bf q}-\omega_{\bf q^{\prime}})
\mu_{{\bf k-q},{\bf q^{\prime}}}\mu_{{\bf k},{\bf q}}+
\mbox{higher order terms}
\\
V^{haa}_{2}({\bf k},{\bf q},{\bf q^{\prime}})=&& 
M_{{\bf k},{\bf q}}\mu_{{\bf k-q-q^{\prime}},{\bf -q^{\prime}}}
 +\mu_{{\bf k},{\bf q}}M_{{\bf k-q-q^{\prime}},{\bf -q^{\prime}}}
-M_{{\bf k-q^{\prime}},{\bf q}}\mu_{\bf k-q^{\prime}, -q^{\prime}}
-\mu_{{\bf k-q^{\prime}},{\bf q}}M_{\bf k-q^{\prime}, -q^{\prime}}
\nonumber \\
&&+\frac{J}{t}(\omega_{\bf q}+\omega_{\bf q^{\prime}})\bigl[
\mu_{{\bf k},{\bf q}}\mu_{{\bf k-q-q^{\prime}},{\bf -q^{\prime}}}
-\mu_{{\bf k-q^{\prime}},{\bf q}}\mu_{\bf k-q^{\prime}, -q^{\prime}}
\bigr]+\mbox{higher order terms}
\nonumber
\end{eqnarray}

\section{}

Hole-two-magnon addition $\delta H_{J}$ (\ref{T7}) originated from the 
projectors applied to $J$-term has the form
\begin{eqnarray}
\label{BB1}
\delta H_{J}=&&J \sum_{{\bf k, k^{\prime}},{\bf Q},{\bf q,q^{\prime}}} 
\left\{V^{J,1}_{{\bf Q},{\bf q},{\bf q^{\prime}}}
\left(h_{\bf k-Q-q-q^{\prime}}^{\dag}
 h_{\bf k^{\prime}+Q}^{\dag} h_{\bf k^{\prime}} h_{\bf k} 
a_{\bf q}^{\dag} a_{\bf q^{\prime}}^{\dag}
+ \mbox{H.c.}\right)  
+ 2 V^{J,2}_{{\bf Q},{\bf q},{\bf q^{\prime}}}
h_{\bf k-Q-q+q^{\prime}}^{\dag}
h_{\bf k^{\prime}+Q}^{\dag} h_{\bf k^{\prime}} h_{\bf k} 
a_{\bf q}^{\dag} a_{\bf q^{\prime}}\right\}\\
&&-J\sum_{{\bf k},{\bf q,q^{\prime}}}\left\{ 
\widetilde{V}^{J,1}_{{\bf q},{\bf q^{\prime}}}
\left(h_{\bf k-q-q^{\prime}}^{\dag} h_{\bf k} 
a_{\bf q}^{\dag} a_{\bf q^{\prime}}^{\dag}
+ \mbox{H.c.}\right)  
+2 \widetilde{V}^{J,2}_{{\bf q},{\bf q^{\prime}}}
h_{\bf k-q+q^{\prime}}^{\dag} h_{\bf k} 
a_{\bf q}^{\dag} a_{\bf q^{\prime}}\right\}\ ,
\nonumber 
\end{eqnarray}
where vertices are given by
\begin{eqnarray}
\label{BB2}
V^{J,1}_{{\bf Q},{\bf q},{\bf q^{\prime}}}=&&
\gamma_{\bf Q}(u_{\bf q}v_{\bf q^{\prime}}+v_{\bf q}u_{\bf q^{\prime}})
+{\textstyle\frac{1}{2}}[\gamma_{\bf Q+q}+\gamma_{\bf Q+q^{\prime}}]
(u_{\bf q}u_{\bf q^{\prime}}+v_{\bf q}v_{\bf q^{\prime}})\ ,
\\
V^{J,2}_{{\bf Q},{\bf q},{\bf q^{\prime}}}=&&
\gamma_{\bf Q}(u_{\bf q}u_{\bf q^{\prime}}+u_{\bf q}u_{\bf q^{\prime}})
+{\textstyle\frac{1}{2}}[\gamma_{\bf Q+q}+\gamma_{\bf Q-q^{\prime}}]
(u_{\bf q}v_{\bf q^{\prime}}+v_{\bf q}u_{\bf q^{\prime}}) \ ,
\nonumber \\
\widetilde{V}^{J,1}_{{\bf q},{\bf q^{\prime}}}=&&
[1+\gamma_{\bf q+q^{\prime}}]
(u_{\bf q}v_{\bf q^{\prime}}+v_{\bf q}u_{\bf q^{\prime}})
+[\gamma_{\bf q}+\gamma_{\bf q^{\prime}}]
(u_{\bf q}u_{\bf q^{\prime}}+v_{\bf q}v_{\bf q^{\prime}}) \ ,
\nonumber \\
\widetilde{V}^{J,2}_{{\bf q},{\bf q^{\prime}}}=&&
[1+\gamma_{\bf q-q^{\prime}}]
(u_{\bf q}u_{\bf q^{\prime}}+v_{\bf q}v_{\bf q^{\prime}})
+[\gamma_{\bf q}+\gamma_{\bf q^{\prime}}]
(u_{\bf q}v_{\bf q^{\prime}}+v_{\bf q}u_{\bf q^{\prime}}) \ ,
\nonumber 
\end{eqnarray}
where $u_{\bf q}$, $v_{\bf q}$ are the parameters of Bogolubov 
transformation, $U_{\bf k}$ defined in (\ref{AA3}).

Applying the transformation (\ref{N6a}) to 
the Hamiltonian  (\ref{BB1}) one can get the additional part to the 
effective interaction (\ref{T7a}) $\delta V^{fg}_{\bf k, k^{\prime},q}
=\delta V^{fg}_{dir}({\bf k, k^{\prime},q})+
\delta V^{fg}_{ex}({\bf k, k^{\prime},q})$ with the direct part
\begin{eqnarray}
\label{BB3}
\delta V^{fg}_{dir}({\bf k, k^{\prime},q})=&&
-\gamma_{\bf q}(1-2\delta \lambda)\bigl[1-U_{\bf k}-U_{\bf k-q}\bigr]\\
&&+4\sum_{\bf q^{\prime}}\mu_{\bf k, q^{\prime}}f_{\bf k-q^{\prime}}
\left\{\delta\lambda\gamma_{\bf k-q^{\prime}}
(\gamma_{\bf q}u_{\bf q^{\prime}}+
{\textstyle\frac{1}{2}}\gamma_{\bf q-q^{\prime}}v_{\bf q^{\prime}})+
{\textstyle\frac{1}{2}}u_{\bf q^{\prime}}
\widetilde{F}_{\bf k-q^{\prime},q}\right\} + 
[{\bf k, q}\Rightarrow {\bf k-q, -q}]
\nonumber \\
&&+4\sum_{\bf q^{\prime}}\mu_{\bf k^{\prime}, q^{\prime}}
f_{\bf k^{\prime}-q^{\prime}}
\left\{{\textstyle\frac{1}{2}}\delta\lambda v_{\bf q^{\prime}}
\gamma_{\bf k^{\prime}-q^{\prime}}\gamma_{\bf q+q^{\prime}}
+{\textstyle\frac{1}{2}}u_{\bf q^{\prime}} 
\widetilde{F}_{\bf k^{\prime}-q^{\prime},-q}
+v_{\bf q^{\prime}}\widetilde{F}_{\bf k^{\prime}-q^{\prime},-q-q^{\prime}}
\right\} + 
[{\bf k^{\prime}, q}\Rightarrow {\bf k^{\prime}+q, -q}]
\nonumber \\
&&-\sum_{\bf q^{\prime}}\mu_{\bf k^{\prime}+q,q-q^{\prime}}
\left\{\mu_{\bf k^{\prime}+q^{\prime},q^{\prime}}
\widetilde{V}^{J,1}_{{\bf q-q^{\prime}},{\bf q^{\prime}}}
+\mu_{\bf k^{\prime},-q^{\prime}}
\widetilde{V}^{J,2}_{{\bf q-q^{\prime}},{\bf -q^{\prime}}}
\right\} + 
[{\bf k^{\prime}, q}\Rightarrow {\bf k^{\prime}+q, -q}] \ ,
\nonumber 
\end{eqnarray}
and exchange part
\begin{eqnarray}
\label{BB4}
\delta V^{fg}_{ex}({\bf k, k^{\prime},q})=&&
32 f_{\bf k}f_{\bf k^{\prime}}
\left\{{\textstyle\frac{1}{2}}\delta\lambda^2\gamma_{\bf k}
\gamma_{\bf k^{\prime}}\gamma_{\bf q}
+\delta\lambda\gamma_{\bf k}\widetilde{F}_{\bf k^{\prime},-q}
+{\textstyle\frac{1}{2}}D_1\gamma_{\bf k^{\prime}}
\widetilde{F}_{\bf k,q}
+{\textstyle\frac{1}{8}}D_2\widetilde{F}_{\bf k,k^{\prime}+q}
+{\textstyle\frac{1}{8}}D_3\widetilde{F}_{\bf k,-k^{\prime}+q}
\right\}\\
&&-2\left\{\mu_{\bf k^{\prime}+q,q}
\Phi^{\it 1}_{\bf k,q}+\Phi^{\it 2}_{\bf k,q}
\right\}
+(1-2\delta\lambda)
\sum_{\bf q^{\prime}}\gamma_{\bf q+q^{\prime}}
\mu_{\bf k^{\prime}, q^{\prime}}\mu_{\bf k+q^{\prime}, q^{\prime}}
\nonumber\\
&&+[{\bf k, k^{\prime}, q}\Rightarrow {\bf k-q, k^{\prime}+q, -q}]\ ,
\nonumber
\end{eqnarray}
where functions
\begin{eqnarray}
\label{BB5}
\widetilde{F}_{\bf k_1,k_2}=&&D_1\gamma_{\bf k_1}\gamma_{\bf k_2}
+{\textstyle\frac{1}{4}}D_2\gamma_{\bf k_1-k_2}
+{\textstyle\frac{1}{4}}D_3\gamma_{\bf k_1+k_2}\ ,\\
\Phi^{\it 1}_{\bf k,q}+\Phi^{\it 2}_{\bf k,q}=&&
4f_{\bf k}\left\{\delta\lambda\gamma_{\bf k}
(\gamma_{\bf q}v_{\bf q}+u_{\bf q})+
({\textstyle\frac{1}{4}}+\delta\lambda)u_{\bf q}\gamma_{\bf k}
+v_{\bf q}\widetilde{F}_{\bf k,q}\right\}+
[{\bf k, q}\Rightarrow {\bf k-q, -q}]\ ,
\nonumber 
\end{eqnarray}
with the numbers
\begin{eqnarray}
\label{BB6}
&&\delta\lambda=\sum_{\bf q}(v_{\bf q}^2+\gamma_{\bf q}v_{\bf q}u_{\bf q})
=-0.0790\ ,
\ \ \ D_1=\sum_{\bf q}(\gamma_{\bf q}v_{\bf q}u_{\bf q}+u_{\bf q}^2
\cos q_x\cos q_y)=-0.1301\ ,
 \\
&&D_2=\sum_{\bf q}u_{\bf q}^2(1-\cos q_x\cos q_y)=0.2628\ ,\ \ \
D_2=\sum_{\bf q}u_{\bf q}^2(2\cos^2 q_x-1-\cos q_x\cos q_y)=-0.0077\ .
\end{eqnarray}

\section{}

For two holes with the total momentum ${\bf P}=0$ one can write the 
following integral equation 
\begin{eqnarray}
\label{T8}
\widetilde{\Gamma}(k_f,-k_g, k_f^{\prime},-k_g^{\prime})=
\Gamma^0(k_f,-k_g, k_f^{\prime},-k_g^{\prime})+
\sum_{p_f}&& 
\Gamma^0(k_f,-k_g, p_f,-p_g) G(p_f) G(-p_g) \widetilde{\Gamma}(p_f,-p_g, 
k_f^{\prime}, -k_g^{\prime})
\end{eqnarray} 
where we introduced four-momentum
notations $k_f=({\bf k},\epsilon_f)$, $-k_g=(-{\bf 
k},\epsilon_g)$, $p_f=({\bf p},\epsilon_f^{\prime\prime})$, $-p_g=(-{\bf 
k},\epsilon_g^{\prime\prime})$
 with momenta ${\bf k, p}$ and frequencies $\epsilon_{f(g)}$, 
$\epsilon_{f(g)}^{\prime\prime}$. $G(p)=1/(\epsilon-E_{\bf k}+i\delta)$ 
is the single-hole Green function.
 This equation is equivalent to the graphical equality shown in 
Fig.  \ref{fig8a}.  Near the pole $\Gamma^0 \ll \widetilde{\Gamma}$ and 
hence the 
first term in Eq.  (\ref{T8}) can be neglected. Then, one can see that 
$\widetilde{\Gamma}$ dependence on outgoing four-momenta
$k$, $k^{\prime}$ is the parametric one, i.e. it is not determined by 
equation, 
and they can be omitted. Let us also introduce 
$E=\epsilon_f+\epsilon_g$,
$\Delta\epsilon=(\epsilon_f-\epsilon_g)/2$, 
and $\Delta\epsilon^{\prime\prime}=(\epsilon_f^{\prime\prime}-
\epsilon_g^{\prime\prime})/2$.
Thus,
\begin{eqnarray}
\label{T9}
\widetilde{\Gamma}({\bf k},E,\Delta\epsilon)=\sum_{{\bf p},
\Delta\epsilon^{\prime\prime}} \Gamma^0({\bf k},{\bf p},E,\Delta\epsilon,
\Delta\epsilon^{\prime\prime}) G({\bf p},E/2+\Delta\epsilon^{\prime\prime}) 
G(-{\bf p},E/2-\Delta\epsilon^{\prime\prime})\widetilde{\Gamma}({\bf 
p},E,\Delta\epsilon^{\prime\prime})
\ . 
\end{eqnarray}
When $\Gamma^0$ has no frequency dependence ("static" interaction), 
\begin{eqnarray}
\label{T10}
\Gamma^0({\bf k},{\bf p},E,\Delta\epsilon,\Delta\epsilon^{\prime\prime})=
U({\bf k},{\bf p}) \ ,
\end{eqnarray}
it is natural to change the variable \  $GG\widetilde{\Gamma}=\chi$ and 
get
\begin{eqnarray}
\label{T11}
\chi({\bf k},E,\Delta\epsilon)=G({\bf k},E/2+\Delta\epsilon) G(-{\bf 
k},E/2-\Delta\epsilon)\sum_{\bf p} U({\bf k},{\bf p})
\int d(\Delta\epsilon^{\prime\prime}) \chi({\bf p},E,
\Delta\epsilon^{\prime\prime}) \ . 
\end{eqnarray}
Additional integration of both sides over $\Delta\epsilon$ gives the 
Schr\"{o}dinger equation:
\begin{eqnarray}
\label{T12}
\psi({\bf k},E)=\frac{1}{E-2E_{\bf k}} \sum_{\bf p} U({\bf k},{\bf 
p}) \psi({\bf p},E) \ ,
\end{eqnarray}
with $\psi({\bf k},E)=\int d(\Delta\epsilon) \chi({\bf 
k},E,\Delta\epsilon)$, which has the sense of the bound-state wave 
function.

In our case the "compact" vertex
$\Gamma^0$ consists of two parts (see Fig. \ref{fig8b}) and one 
has to include the magnon propagator into expression for 
the spin-wave exchange vertex
\begin{eqnarray}
\label{T13}
\Gamma^0_1({\bf k},{\bf p},\Delta\epsilon,\Delta\epsilon^{\prime\prime})=
-\biggl[ \frac{V_{{\bf k},{\bf p}} V^*_{-{\bf k},-{\bf p}}
}{\epsilon-\epsilon^{\prime\prime}-\omega_{{\bf k}+{\bf p}}+i\delta}+
\frac{V_{-{\bf k},
-{\bf p}} V^*_{{\bf k},{\bf p}} }{\epsilon^{\prime\prime}-\epsilon-
\omega_{-{\bf k}-{\bf p}}+i\delta} \biggr] 
\end{eqnarray}
where $V_{\bf k, p}$ is the vertex of the type Eq. (\ref{T2}),
$\epsilon-\epsilon^{\prime\prime}=\Delta\epsilon-
\Delta\epsilon^{\prime\prime}$, ${\bf k}+{\bf p}={\bf q}$.
Both negative sign and ${\bf q}={\bf k}+{\bf p}$ are due to exchange 
character of the diagram (Fig. \ref{fig8b}). 
Thus, there are three $\Delta\epsilon^{\prime\prime}$-dependent 
denominators in the 
integral Eq. (\ref{T9}) and the above change of variables is impossible.

It is natural to assume that since one is looking for the poles of the
two-particle Green function as the function of $E$, 
$\widetilde{\Gamma}$ has no singularities as the function of 
the difference 
of the energies of incoming particles. 
Therefore, the integration over 
$\Delta\epsilon^{\prime\prime}$ in Eq. (\ref{T9}) can be done and it 
is determined by the 
poles of $G({\bf p},E/2+\Delta\epsilon^{\prime\prime})$, $G({\bf 
p},E/2-\Delta\epsilon^{\prime\prime})$, and $\Gamma^0_1({\bf k},{\bf 
p},\Delta\epsilon,\Delta\epsilon^{\prime\prime})$ (\ref{T13}). 
Their poles are:  
$\Delta\epsilon^{\prime\prime}=(E_{\bf p}-E/2)-i\delta$,  
$\Delta\epsilon^{\prime\prime}=-(E_{\bf p}-E/2)+i\delta$, and  
$\Delta\epsilon^{\prime\prime}=\pm (\Delta\epsilon-\omega_{\bf q})\pm 
i\delta$, 
respectively. $+ (-)$ in the last pole corresponds to the first (second) 
term in Eq. (\ref{T13}). The integration gives
\begin{eqnarray}
\label{T14}
\widetilde{\Gamma}({\bf k},E,\Delta\epsilon)=\sum_{\bf p}
\left(-\frac{V_{{\bf k},{\bf p}} V^*_{-{\bf k},-{\bf p}}
}{E-2E_{\bf p}}\right)\left[ 
\frac{\widetilde{\Gamma}\left({\bf p},E,(E_{\bf p}-E/2)\right)
}{\Delta\epsilon-(E_{\bf p}-E/2)-\omega_{\bf q}+i\delta}+
\frac{\widetilde{\Gamma}\left({\bf p},E,-(E_{\bf p}-E/2)\right)
}{-\Delta\epsilon-(E_{\bf p}-E/2)-\omega_{\bf q}+i\delta} \right] \ ,
\end{eqnarray}
The further way is close to the usual one. Multiplying both sides 
of Eq. (\ref{T14}) by the 
external incoming Green functions one can integrate over $\Delta\epsilon$, 
using the evident parity of $\widetilde{\Gamma}$ on $\Delta\epsilon$
\begin{eqnarray}
\label{T15}
\frac{\widetilde{\Gamma}({\bf k},E)}{E-2E_{\bf k}}
=\frac{1}{E-2E_{\bf k}}
\sum_{\bf p}
\frac{-2V_{{\bf k},{\bf p}} V^*_{-{\bf k},-{\bf p}}       
}{E-E_{\bf p}-E_{\bf k}-\omega_{\bf q}}\times
\frac{\widetilde{\Gamma}({\bf p},E)}{E-2E_{\bf p}}\ .
\end{eqnarray}
 Changing $\widetilde{\Gamma}({\bf 
k},E)/(E-2E_{\bf k})=\psi({\bf k},E)$ one obtains
\begin{eqnarray}
\label{T16}
\psi({\bf k},E)=\frac{1}{E-2E_{\bf k}} \sum_{\bf p} 
\frac{-2V_{{\bf k},{\bf p}} V^*_{-{\bf k},-{\bf p}}
}{E-E_{\bf p}-E_{\bf k}-\omega_{\bf q}}
\psi({\bf p},E) \ .
\end{eqnarray}
Evidently, the "usual" Bethe-Salpeter equation (\ref{T12})
can be obtained by the same way. 
Surprisingly, this result (\ref{T16}) coincides exactly with one 
obtained in Ref. \cite{SK} using Rayleigh-Schr\"{o}dinger perturbation 
theory.


\vskip 1cm
{\Large \bf Figure captions} \\ \vskip 0.1cm
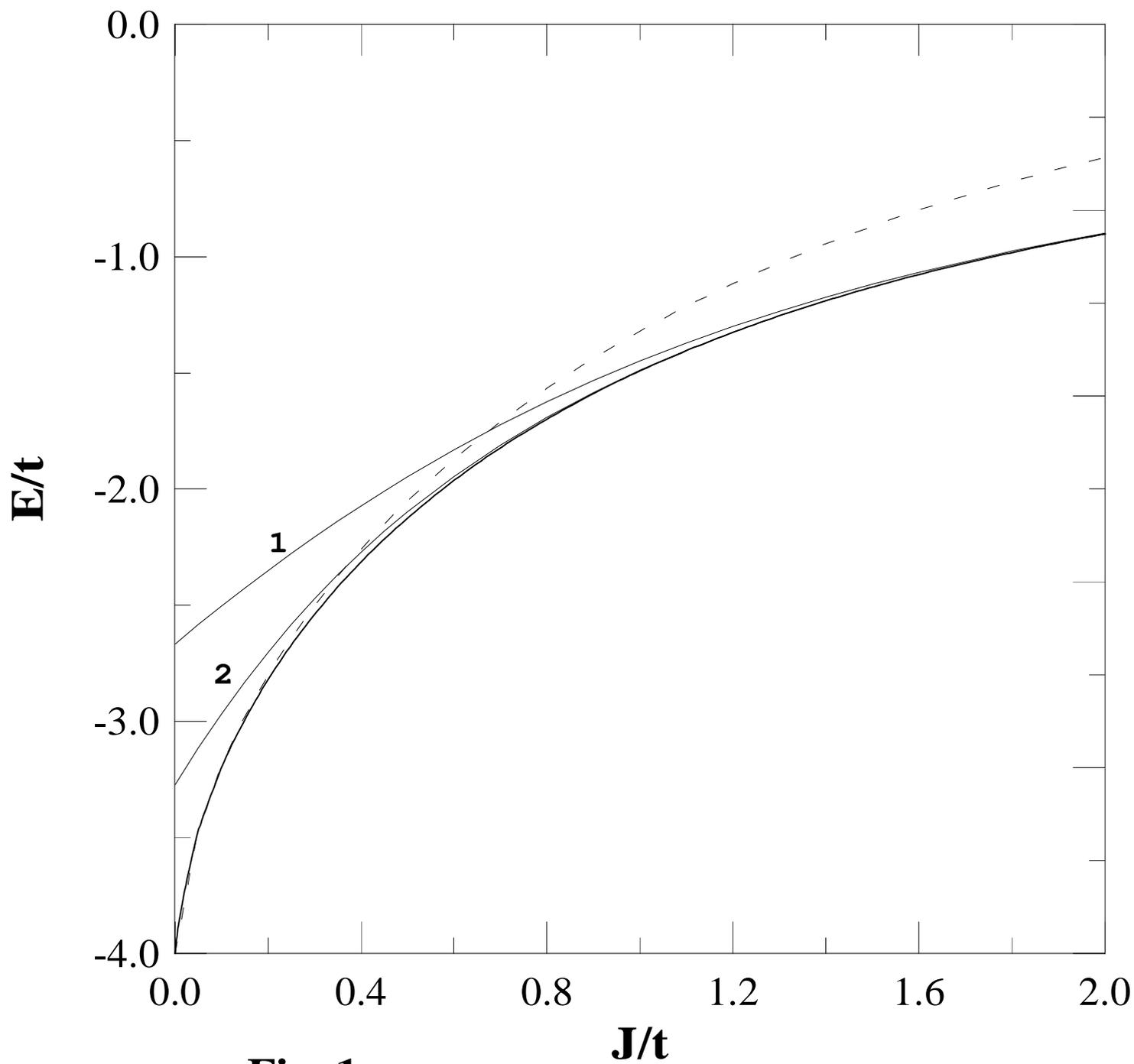
\begin{figure}
\caption{ Single-hole energy for the Ising limit. Bold solid curve is an 
exact result in the spin-wave approximation. Dashed curve is an exact result 
for the large $t\gg J$ limit. Solid curves (1) and (2)  are the canonical 
transformation results up to the fourth and sixth order of transformation, 
respectively.
 } \label{fig1} 
\end{figure} 
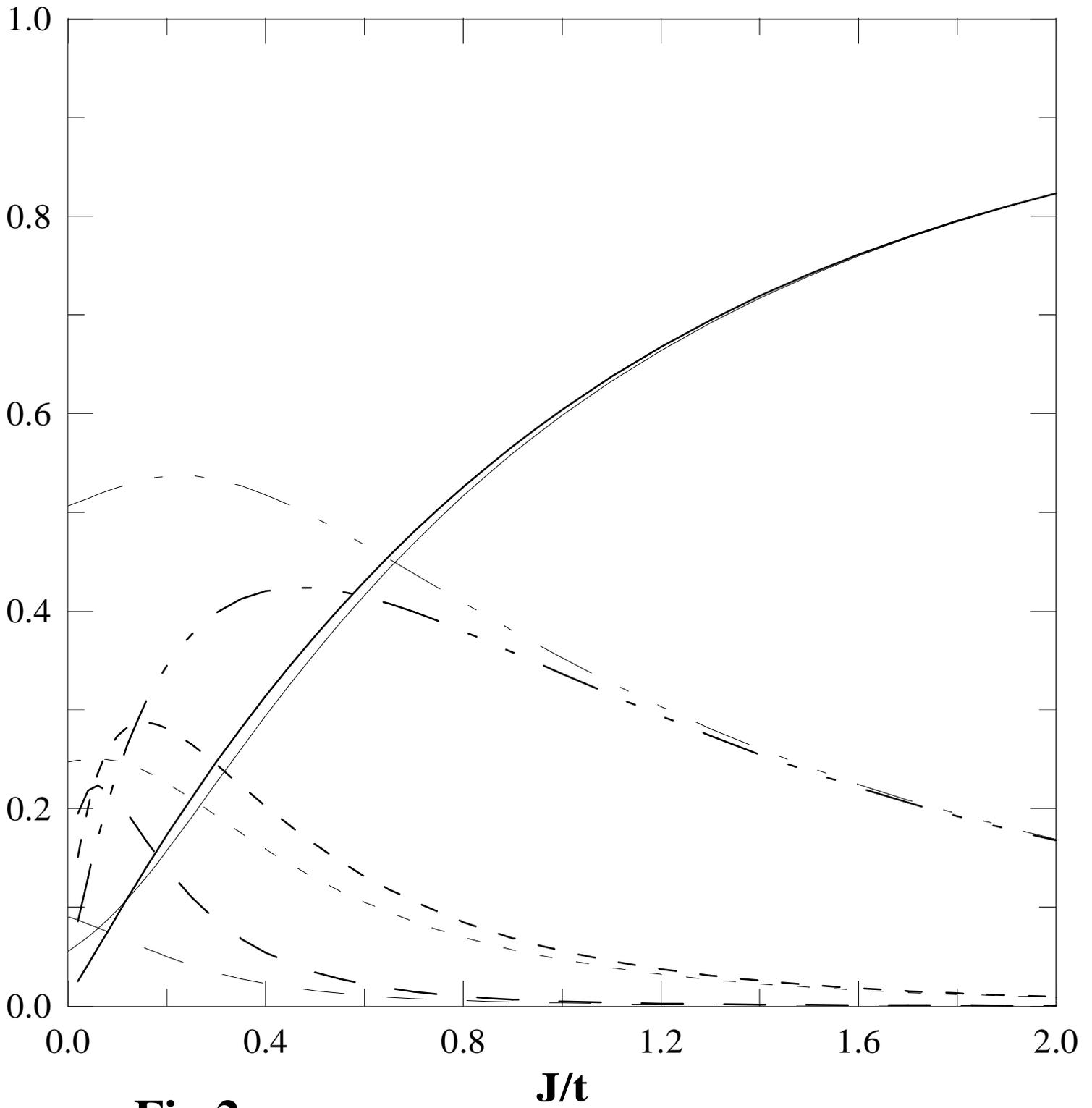
\begin{figure}
\caption{ Weights of the components of the magnetic polaron 
 wave function (Ising limit). Bold 
curves are for an exact solution, light curves are for the canonical 
transformation results (fifth order). Solid curves correspond to the weights 
of the bare hole, dashed-double-dotted curves -- hole$+$1 magnon, dashed 
curve -- hole$+$2 magnons, long-dashed curve -- hole$+$3 magnons.  } 
\label{fig2}
\end{figure}
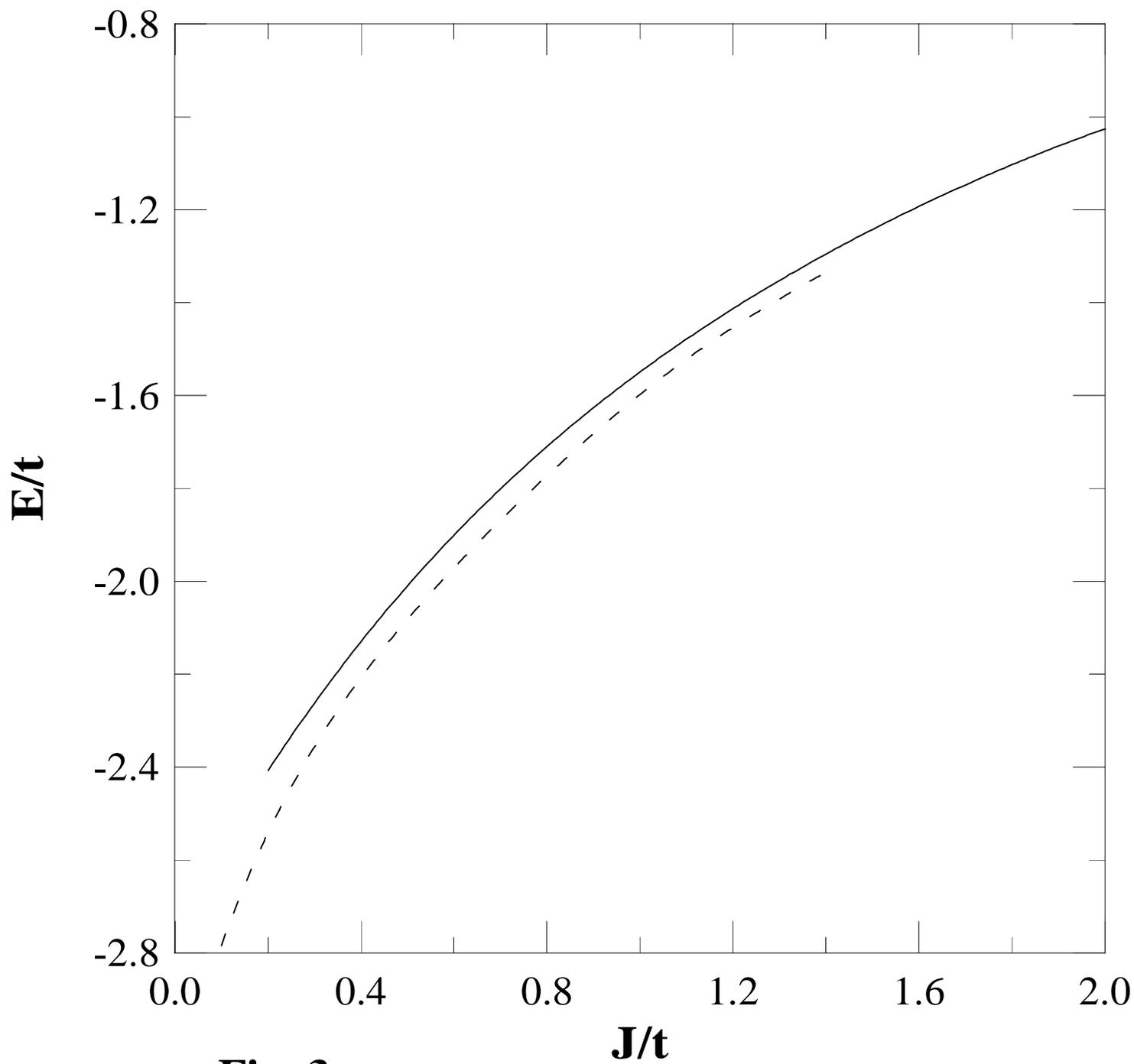
\begin{figure}
\caption{ Bottom of the hole band. Solid curve is our result (the sixth 
order of the transformation), dashed curve is the SCBA result.  } 
\label{fig3}
\end{figure}
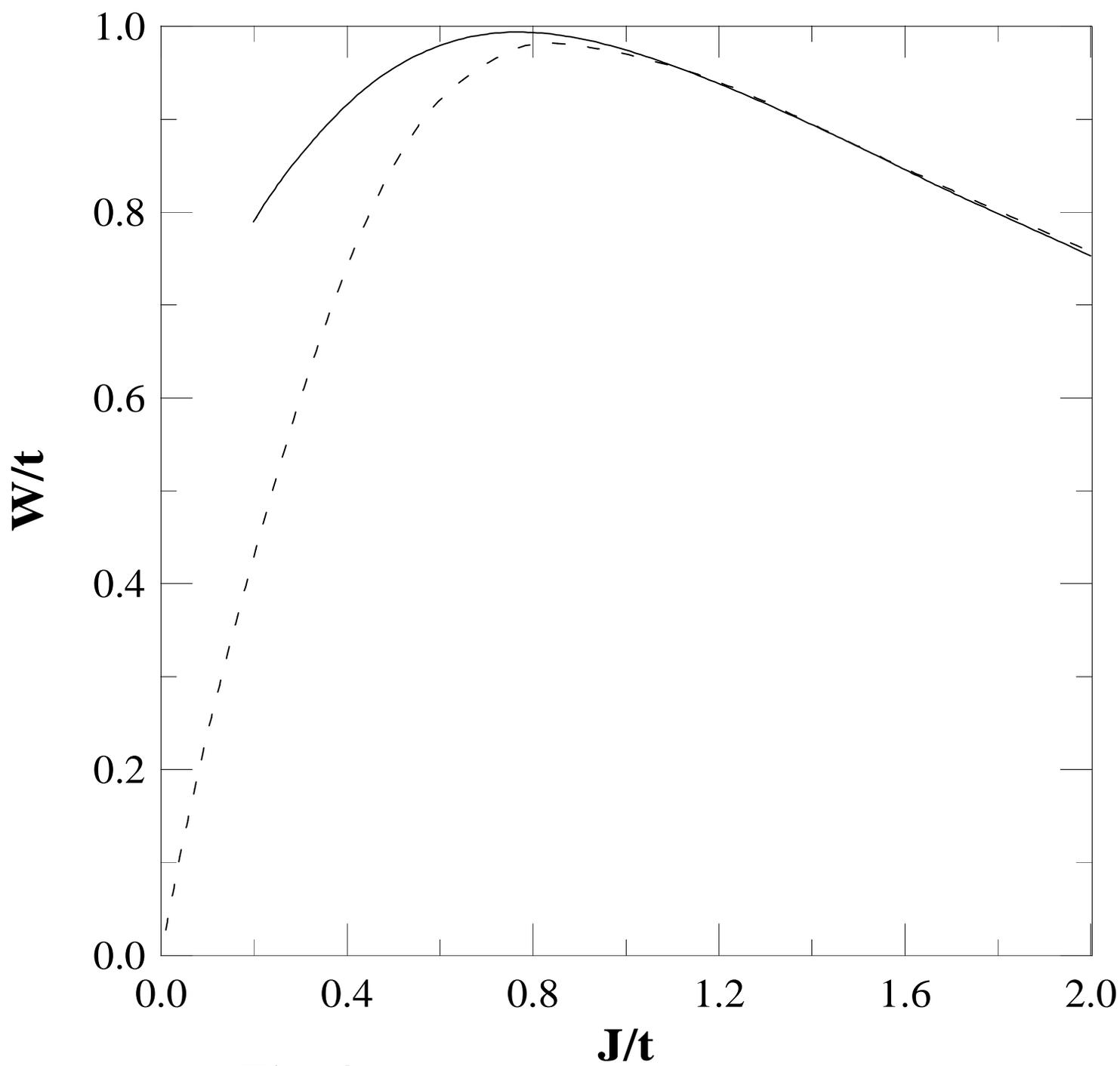
\begin{figure}
\caption{ Width of the hole band. Solid curve is our result (the 
sixth order of the transformation), dashed curve is the SCBA result.
} 
\label{fig4}
\end{figure}
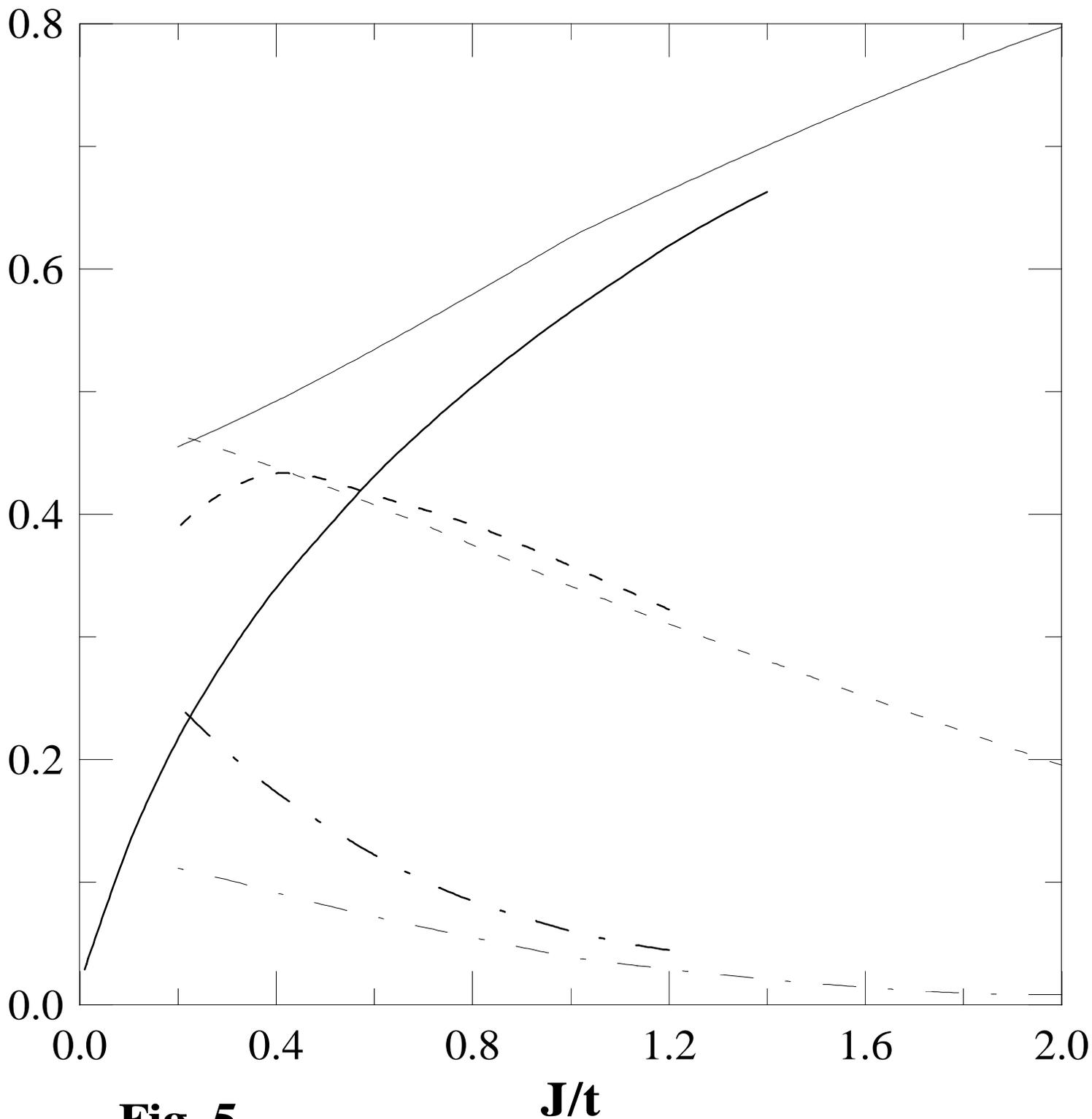
\begin{figure}
\caption{ Weights of the components of the magnetic polaron 
 wave function (N\'{e}el case). Bold curves 
are for an exact solution (SCBA), light curves are for the canonical 
transformation results (the fourth order). Solid curves correspond to the 
weights of the bare  hole, dashed curves -- hole$+$1 magnon, dashed-dotted 
curve -- hole$+$2 magnons.
} 
\label{fig5}
\end{figure}
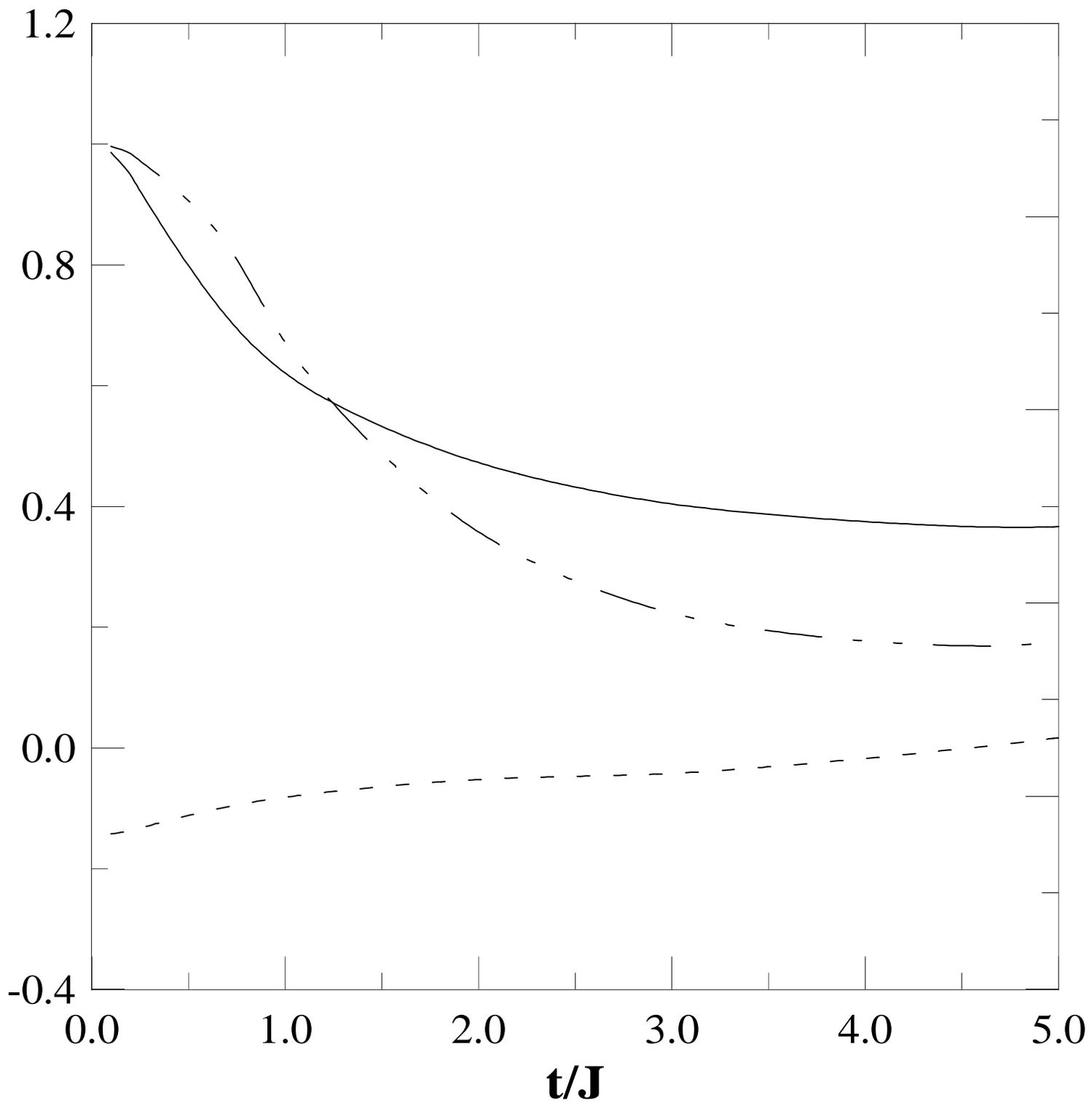
\begin{figure}
\caption{  $t/J$ dependence of the formfactor $F_{\bf k,q}$.
Solid curve -- ${\bf k}=(\pi/2,\pi/2)$, ${\bf q}=(0,0)$,
dashed curve -- ${\bf k}=(\pi/2,\pi/2)$, ${\bf q}=(\pi/2,\pi/2)$,
 dashed-double-dotted curve -- ${\bf k}=(0,0)$, ${\bf q}=(0,0)$.
}               
\label{fig6}
\end{figure}
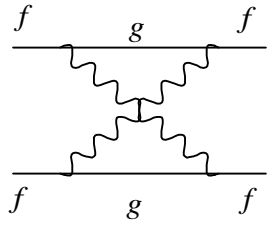
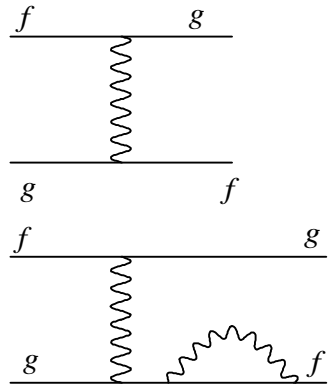
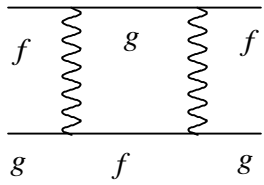
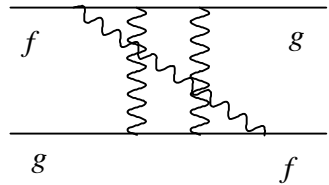
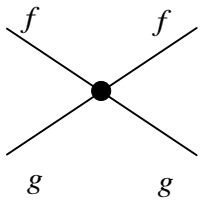
\begin{figure}
\caption{  Schematic view of the scattering diagrams: (a) 
$ff\rightarrow ff$, (b) $fg\rightarrow gf$ of exchange type,
(c) $fg\rightarrow fg$ of direct type, (d) the first diagram of the
$fg\rightarrow fg$ scattering beyond the ladder approximation. 
Here, the wavy lines denote the interaction, which originates from the 
magnon exchange ($t$-term), and 
the point in the diagram (c) denotes the vertex, which is from the 
nearest neighbor interaction ($J$-term).  
} 
\label{fig7} 
\end{figure} 
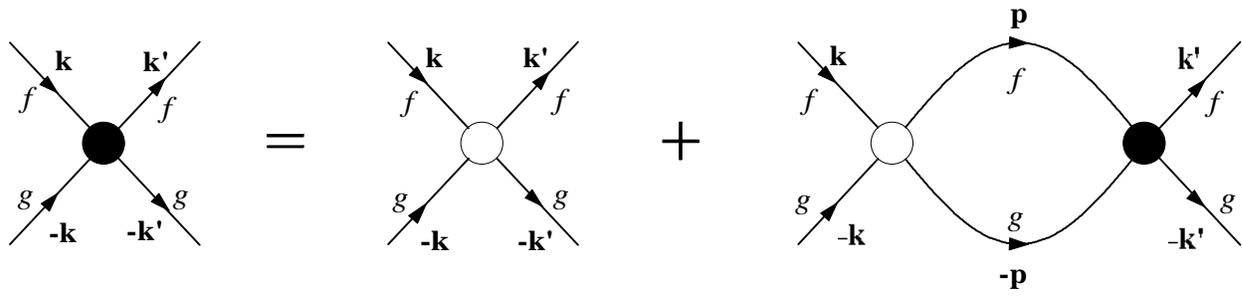
\begin{figure}
\caption{  Graphical identity for an exact vertex  
$\widetilde{\Gamma}(k_f,-k_g, k_f^{\prime},-k_g^{\prime})$ for the $fg$
scattering in the ladder approximation. Black circle denotes 
$\widetilde{\Gamma}(k_f,-k_g, k_f^{\prime},-k_g^{\prime})$, empty 
circle denotes a "compact" vertex 
$\Gamma^0(k_f,-k_g, k_f^{\prime},-k_g^{\prime})$.
} 
\label{fig8a}
\end{figure}
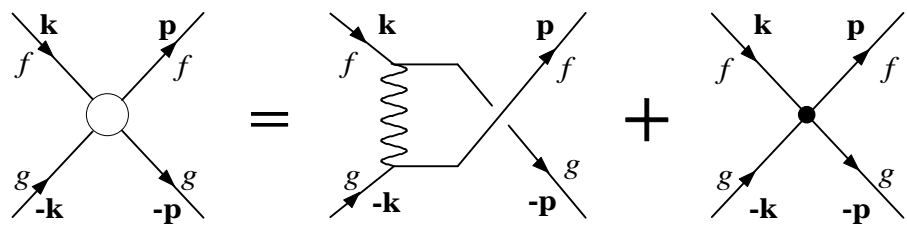
\begin{figure}
\caption{ Structure of the "compact" vertex $\Gamma^0(k_f,-k_g, 
k_f^{\prime},-k_g^{\prime})$ (empty circle). Here,
the wavy line denotes the interaction due to the long-range spin-wave 
exchange (\protect\ref{T2}), and
the point denotes all diagrams, which do not contain the retardation 
(\protect\ref{T4}), (\protect\ref{T7a}).
} 
\label{fig8b}
\end{figure}         
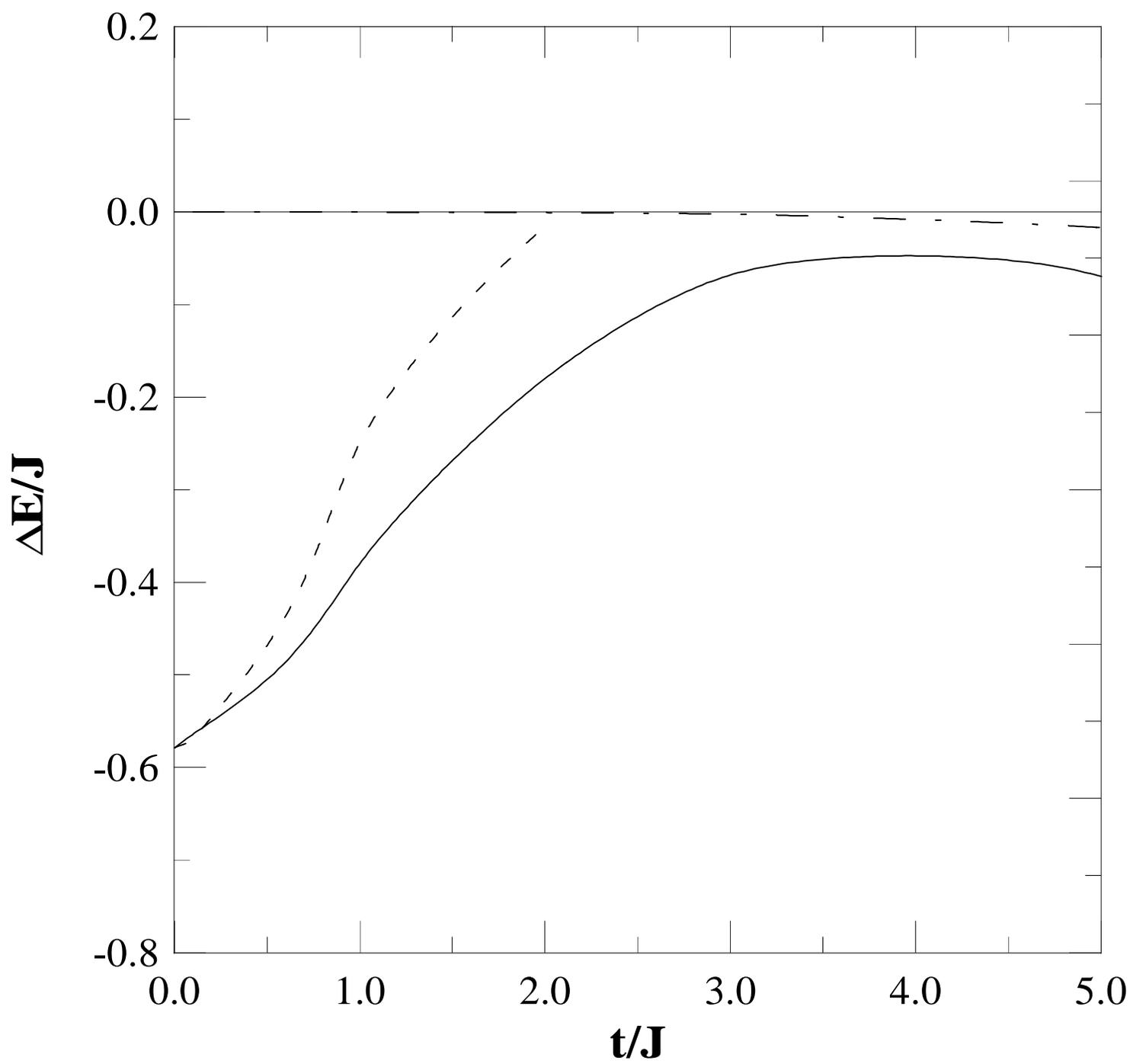
\begin{figure}
\caption{  Results for the energy of the $d$-wave pairing state.
Dashed curve corresponds the short-range bound state,
dashed-dotted curve corresponds to the long-range one, and
solid curve corresponds to the combined effect of both types of the state.
} 
\label{fig9}
\end{figure}
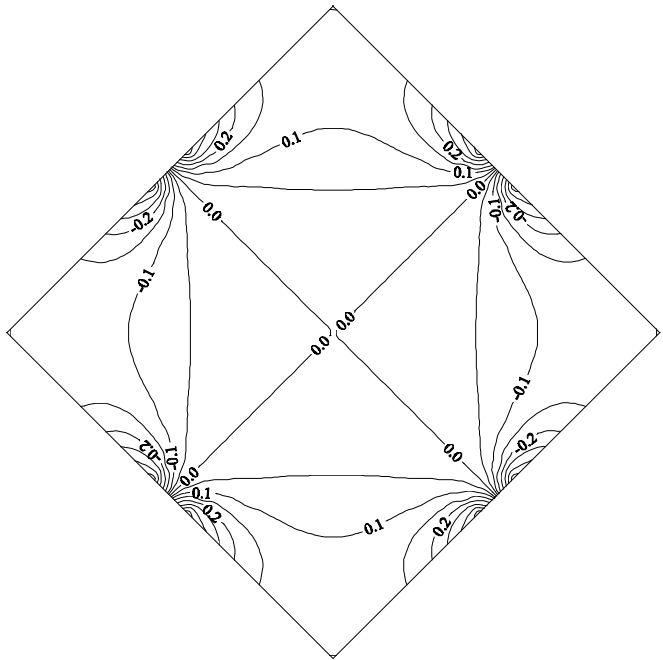
\begin{figure}
\caption{ Wave functions of the two-hole bound states:
(a) $(t/J)=1$, (b) $(t/J)=2$, long-range state (c) $(t/J)=2$, short-range 
one, (d) $(t/J)=2$, wave function of the mixed state.  
} 
\label{fig10}
\end{figure}


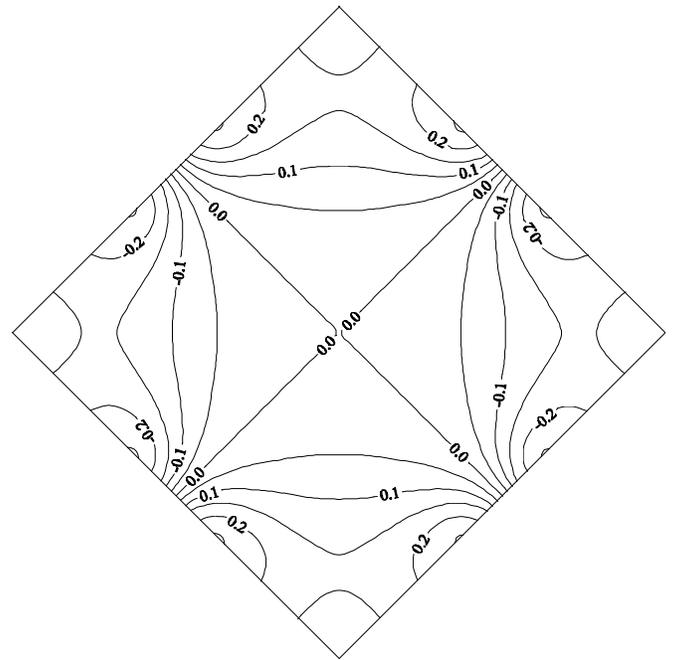
\end{document}